\documentclass[3p, final, authoryear]{elsarticle}
\usepackage{tgtermes} 
\usepackage{graphicx} 
\usepackage{enumitem}
\usepackage{subcaption}%
\usepackage{float}%
\usepackage{makecell}

\usepackage{geometry}

\usepackage{tocloft}
\usepackage{xcolor, soul}
\sethlcolor{yellow}

\usepackage{xcolor}
\usepackage{fancyhdr}

\begin{document}

\begin{frontmatter}

\journal{arXiv}

\title{Statistical arbitrage in multi-pair trading strategy based on graph clustering algorithms in US equities market}


\author[1]{Adam Korniejczuk\corref{cor1}%
\fnref{fn1}}
\ead{a.korniejczu@student.uw.edu.pl}

\author[2]{Robert Ślepaczuk\fnref{fn3}}
\ead{rslepaczuk@wne.uw.edu.pl}

\cortext[cor1]{Corresponding author: a.korniejczu@student.uw.edu.pl}

\fntext[fn3]{ORCID: 0000-0001-5227-2014}

\affiliation[1]{
    organization={Faculty of Economic Sciences, University of Warsaw},
    addressline={ul. Długa 44/50},
    postcode={00-241},
    city={Warsaw},
    country={Poland}
    }
\affiliation[2]{
    organization={Quantitative Finance Research Group, Department of Quantitative Finance, Faculty of Economic Sciences, University of Warsaw},
    addressline={ul. Długa 44/50},
    postcode={00-241},
    city={Warsaw},
    country={Poland}
    }

\date{May 2024}

\begin{abstract}
The study seeks to develop an effective strategy based on the novel framework of statistical arbitrage based on graph clustering algorithms. The amalgamation of quantitative and machine learning methods, including the Kelly criterion, and an ensemble of machine learning classifiers have been used to improve risk-adjusted returns and increase the immunity to transaction costs over existing approaches.  The study seeks to provide an integrated approach to optimal signal detection and risk management. As a part of this approach, innovative ways of optimizing take profit and stop loss functions for daily frequency trading strategies have been proposed and tested. All of the tested approaches outperformed appropriate benchmarks. The best combinations of the techniques and parameters demonstrated significantly better performance metrics than the relevant benchmarks. The results have been obtained under the assumption of realistic transaction costs, but are sensitive to changes in some key parameters.
\end{abstract}

\end{frontmatter}

\section{Introduction}

\par This research aims to develop an effective algorithmic trading strategy, which would be built upon the novel framework involving graph clustering algorithms for statistical arbitrage. The aim is important as the study contributes to the growing body of research exploring the applications of graph theory in the algorithmic trading context. Furthermore, the detailed analysis of the strategy performance will allow for the accurate assessment of the strengths and weaknesses of the novel framework. In this study, the machine learning classifiers are utilized to enhance the average profitability of the trades executed. Moreover, throughout the implementation of the strategy, numerous optimizations are tested. Thus four research questions, around which the study is organized have been formulated:
\begin{itemize}
\item RQ1: Does the usage of signal quality classifiers improve the quality of the graph-clustering-based strategy?
\item RQ2: To what extent, does the implementation of transaction and risk management measures to influence the performance of the strategy?
\item RQ3: What is the sensitivity of the strategy to changes in transaction costs?
\item RQ4: To what extent does the change of weights in the classifiers ensemble influence the strategy performance?
\end{itemize}
\par  The study is based on the daily data of constituents of S\&P500, spanning from January 1, 2000, to December 31, 2022. To address the formulated questions, the proposed approach involves creating an ensemble classifier of the quality of signals generated by a graph-based strategy. Individual classifiers are trained and optimized using grid search cross-validation, then combined in the weighted ensemble based on their performance on the validation dataset. Finally, the strategy involving classifiers, Kelly criterion, and innovative stop loss and take profit functions is tested on out of sample, testing dataset.
\par Several novelties are present in the study. Firstly, the study aims to implement machine learning classifiers to filter the potentially profitable signals generated by another strategy. As far as is known, this approach has not been implemented previously in the context of algorithmic trading strategy, as usually the classification models are used for directional changes of stock prices. Moreover, new, time-variant stop loss and take profit functions are introduced and tested with the study. Furthermore, the modification of those functions is combined with signal-strength scaling of take profit and stop loss functions, resulting in an unconventional approach. The research examines the potential of emerging and almost unexplored graph-based approach. It makes it highly relevant, as technological advancements increased the competitiveness in the field of quantitative trading over the years, thus making it more difficult to sustain stable abnormal returns over a prolonged time and creating the need for new, sophisticated methods. 
\par The structure of this paper is as follows: After an introduction in the first section,  the second section includes the literature review, giving a theoretical introduction to graph theory and its applications in the context of algorithmic trading as well as examining the current research regarding tools and techniques that are planned to be included in the strategy development. The third section describes the methodology, including the data used, motivation for the chosen approach, feature selection, engineering, the classifiers ensemble building details, and the proposed decision and risk management optimizations. The fourth section presents all the metrics used in evaluating the strategies, then evaluates the proposed strategy's performance using those metrics against the relevant benchmarks and displays their equity curves. In the fifth section sensitivity analysis has been conducted, to measure the impact of change of key parameters on model performance. Finally, the sixth section contains the summary of the results obtained, conclusions, and suggestions for possible further research.
\section{Literature Review}
\par Utilizing information about the relationship between stocks in the construction of investment portfolios or algorithmic strategies has a long history. The rationale behind the approach is straightforward. Companies and subsequently their stocks' prices do not exist in a vacuum, thus, assuming not the perfect market's efficiency, information regarding one stock may be useful when trading other, related stocks. A very influential and early work in the area was written by Markowitz (1952), who formed a base of Modern Portfolio Theory. In this framework, the correlation is used together with variance to find the optimal portfolio, which would be characterized by maximum returns for a given amount of risk. 
\par Since Markowitz's times, various investment strategies leveraging relationships between assets have emerged. The simplest approach might be grouping stocks by economy sector and generalizing the behavior of stocks from certain sectors. For example, Fama and French (1997) grouped stocks into 48 groups based on standardized industrial classification codes, to further examine their properties.
\par More advanced strategies leveraging information about stock relationships include pair trading or statistical arbitrage. In those strategies, one intends to capitalize on short-term deviation from some established relationship between asset prices. Different measures have been applied to identify pairs or groups of mutually connected assets in the context of pair trading or arbitrage strategy, including cointegration e.g. (Kim, 2011), correlation e.g.(Chen et al., 2019), euclidean distance e.g. (Gatev et al., 2006), Hurst exponent (Bui and Ślepaczuk, 2022) or some combination of those e.g. (Miao, 2014). However, results obtained by Bredthauer and Stübinger (2017) who tested multiple combinations of pair trading approaches found in the literature, indicate, that most of those strategies displayed worsening performance in recent times, most likely due to the increasing proliferation of such techniques. Such findings are also supported by Do and Faff (2012) and Gatev et al. (2006), implying that it becomes harder to capitalize on arbitrage opportunities, thus creating a need for novel, more sophisticated approaches. 
\par Among different pairwise relationship-based trading strategies, the area of research that displays a considerable amount of growth and success is a graph-based approach to quantitative trading. Graphs are structures studied in discrete mathematics, which are a very convenient tool for representing relationships between objects. They are composed of nodes, also referred to as vertices, and connections between them, called edges. Two types of graphs can be considered:
Directed Graph is an ordered pair $(V, E)$ such that $V$ is a finite set of vertices and $E$ is a subset of $V \times V$. Elements of $E$ are referred to as edges. An undirected Graph is an ordered pair $(V, E)$ such that $V$ is a finite set of vertices and $E$, a set of edges is a subset of a set of all unordered pairs from $V$, that is of a set $ \{(u,v): u \neq v, u, v \in  V\}$ (Zawadowski 2020). One can also distinguish between weighted and unweighted graphs, where in the case of the former, edges have an assigned weight, which informs about the strength of a relationship. Similarly, graphs could be signed, which means that edges can represent both positive and negative relationships between nodes. An important property of a vertex in a graph is its degree, which is the sum of all edges, having one end in that vertex. A commonly used form of graph representation involves the use of an adjacency matrix. In the case of an undirected graph, a $I \times J$ symmetrical matrix $A$ stores information about the edge connecting vertices $m$ and $n$ in the $A_{m,n}$, and $A_{m,n}$. Such representation is congruent with the structure of the correlation matrix, thus by using the correlation matrix of (for example) returns one can analyze a group of assets using the graph-based tools. As mentioned, a growing number of scholars apply graph theory-based techniques to the investing and algorithmic trading realm.
\par Zhan et al. (2015) Used the neighbor net-based techniques to demonstrate, that the use of correlation-based clustering is an effective way of risk reduction in investment portfolio creation. Li et al. (2022) Represented a heterogeneous dataset containing news texts, graphical indicators, and transaction data in the form of a weighted graph. The graph was later used in the training of graph neural network, to predict the stock market volatility. The accuracy of the results obtained was more than 14\% higher than that of similar methods.
\par Similarly, Cheng et al. (2022) applied multi-modality graph neural networks for time series forecasting. The model has been trained on a dataset combining news with market data, and importantly included both directed and undirected edges, it predicted next-day price movement based on the last 60 trading days. The approach outperformed other state-of-the-art benchmarks, including LSTM-based approaches.
\par Yin et al. (2022 ) used a novel machine learning model called Graph Attention Long Short Memory (GALSTM) for stock price predictions. The recent graph neural network was intended to learn the patterns of correlations between stocks' prices over time. The authors managed to achieve outstanding results of annual returns and daily standard deviation equal to 44.71\%  and 0.42\% respectively. However, it needs to be noted, that the trading system presented in the study was to a great extent optimized for the optimal performance in the Chinese A-share market, and the testing period has been relatively short as it spanned only 13 weeks.
\par An important and novel study in the application of graph techniques to algorithmic trading was published in September 2023 by Alvaro Cartea, Mihai Cucuringu, and Qi Jin. The paper, titled: "Correlation Matrix Clustering for Statistical Arbitrage Portfolios" is the first one to apply graph clustering algorithms to construct statistical arbitrage portfolios. 
In the study, stocks from the US market have been represented as vertices of undirected, signed, weighted graph. Meanwhile, the correlations between the residual returns from the last sixty days of those stocks have been used as weights for the edges. Thus, the correlation matrix of the stocks considered has been the adjacency matrix of the constructed graph. Such a representation allowed for the usage of graph clustering algorithms. A few clustering algorithms have been considered, including Spectral clustering (Ng et al., 2002), two versions of Signed Laplacian Clustering (Kunegis et al., 2010), and the two versions of Signed Positive Over Negative Generalized Eigenproblem (SPONGE) clustering  (Cucuringu et al., 2019) algorithms. Moreover, three techniques for determining the optimal number of clusters have been tested. The first one is keeping the fixed numbers of 30 clusters. In the second approach, the number of clusters corresponds to the smallest number of eigenvectors which explain 90\% of the variance of the correlation matrix upon which clustering is performed. The third approach makes use of the asymptotic properties of Marchenko-Pastur distribution and random matrix theory.
\par Out of those choices, variants utilizing the second or third approach to finding the optimal number of clusters as well as the symmetrical version of the SPONGE algorithm, referred to as SPONGE$_{sym}$ performed the best. As described by Cartea et al. (2023) the SPONGE$_{sym}$ algorithm decomposes the adjacency matrix $A$ into $A^+$ and $A^-$, containing only positive and negative entries respectively. Then it computes the Laplacian matrices $L^+$ and $L^-$, each calculated by the formulas $L^+ = A^+ - D^+$ and $L^- = A^- - D^-$, where $D$ is the diagonal matrix based on the matrix $A$, representing the degree of each vertex. In the next step $L^+_{sym} $, given by: $(D^+)^{-\frac{1}{2}}L^+(D^+)^{-\frac{1}{2}}$ is calculated ( $L^-_{sym}$ is also computed analogously). Then, the algorithm finds k smallest generalized eigenvectors of $(L^+_{sym} +\tau^-I, L^-_{sym} +\tau^+I)$. Finally, the k-means++ algorithm is used to find clusters in the Induced K-dimensional Euclidean space.
\par The innovative approach of the authors of the study led to significant results. The strategy utilizing SPONGE$_{sym}$ outperformed the strategy using fixed FamaFrench sector-based clusters and significantly outperformed the buy and hold benchmark, achieving annualized returns equal to 12.2\%, Sharpe ratio of 1.1 and Sortino ratio of 2.01. However, the overall implementation of the statistical arbitrage strategy seems to be simple, which most likely has been dictated by the focus of the study - testing a novel graph clustering approach. Precisely, upon performing the clustering, the mean of the last five days' returns have been calculated for each cluster. Then long and short positions were opened on assets whose returns during the last five days were lower or higher, respectively than the calculated clusters' means. Such a procedure was repeated every three days. No attempts at optimizing the strategy in regards to optimal allocation, beyond the choice of clustering algorithm have been mentioned in the study. Moreover, no information on the transaction costs being considered was present in the paper. Thus, an attempt at optimizing the strategy presented, and further inspecting the potential of graph clustering algorithms in quantitative trading, will be the main focus of this study.
\par There is a significant body of literature applying machine learning classification algorithms to predict an asset's price change direction. Older studies by Yangru and Zhang (1997) and by Chen et al. (2000) suggest that the prediction of the direction of price change may be a more effective tool for trading strategy, than predicting the value of price indices.  Moreover, the presence of studies achieving very good results (Borovkova, 2019; Dixon, 2017;  Kryńska and Ślepaczuk, 2023) signifies the validity of such an approach.  Furthermore, Bieganowski and Ślepaczuk (2024), demonstrated the validity of a classification-based approach over regression one, in the context of high-frequency data. In their study, the performance of the forecasting base strategy has been examined in four configurations, with varying levels of complexity. In order to rigorously assess the potential of the introduced strategy, it has been tested on cryptocurrency, stocks, and currency prices. Importantly, the best results have been observed in the case of an approach combining supervised autoencoders, combined with triple barrier labeling.
\par Furthermore, Worasucheep (2021) showcased, that the usage of an ensemble of machine learning classifiers can lead to better performance than the use of one model. In his study, the author combined artificial neural network, support vector machine, random forest, extreme gradient boosting, and light gradient boosting. The results demonstrated, that the ensemble outperformed both buy and hold benchmark as well as strategies utilizing individual classifiers, in terms of both returns and Sharpe ratio. Similarly, Zelenkov (2017) has managed to outperform various benchmarks by utilizing the ensemble with weighted voting, in which the weights were optimized using the genetic algorithm. 
\par In spite of the described success in using the classification algorithms in an algorithmic trading context, the approach based on using machine learning algorithms classifiers for improving the quality of some strategy's signals seems to be novel and unexplored. A hypothetical advantage of such a method could be the possibility of using risk-weighted decision-making and portfolio construction techniques, by utilizing probability estimates returned from the model. An example of such a technique is the Kelly criterion. 
\par Kelly criterion (Kelly, 1956) is a formula for maximizing the expected value of the logarithm of wealth. Kelly has shown, that under sufficient assumption, the capital managed with the formula will, with a probability equal to one surpass the capital managed in other ways. Moreover, the author of the formula pointed out, that even though the example of gambling has been given in the derivation, the concept can be applied to any economic situation, where one can operate on fractions of the total capital and reinvesting is possible. Since then Thorp (1969) proposed the usage of the Kelly criterion in various games, such as Blackjack, Roulette, or The Wheel of Fortune, but also in Stock Market investing. Evidence exists (Anderson and Faff, 2004; Bronakowska et al. 2021; Hagman, 2014), suggesting that using the Kelly criterion can improve the profitability of trading strategies. Furthermore, what is important in the context of this study, it has been also successfully applied to algorithmic trading strategies (Chen et al., 2021; Li et al. 2020). At the same time, contrary evidence can also be found (Lähteenmäki, 2023), suggesting that the usage of the formula may not be optimal in all of the cases. 
\par In conclusion, leveraging relationships between stocks is a classical tool in the algorithmic trading strategy domain. Strategies like statistical arbitrage or pair trading, based on measures like correlation or cointegration have become more widespread, thus making them less profitable. Studies from an emerging branch of research involving graph theory-based approaches have been characterized by very good performance, often outperforming other state-of-the-art methods. However, good performance is often linked to a very complicated architecture, involving complex implementations of graph neural networks, or graphs based on multi-modal datasets. Nevertheless, a recent study by Cartea et al. (2023) reveals a new framework involving graph clustering, which can potentially outperform traditional approaches in statistical arbitrage applications. The innovative approach definitely has a lot of potential thus it will serve as the base of this study. The proposed solution of filtering the signals based on their quality seems to be novel, as classifiers have been commonly and successfully used for some kind of buy/sell problems. Although applied in slightly different settings, ensemble classifiers showed promising results, thus such an approach will be utilized in this research. Moreover, the long history and successful applications of Kelly criterion in the algorithmic trading context, are convincing arguments to optimize the strategy using this tool.

\section{Methodology}
\subsection{Data used}
The data used in the study comes from the Yahoo Finance database. In order to ensure a sufficient degree of realism, historical components of the S\&P 500 index have been used, such that during backtesting, positions could be opened only on the stocks which were historically present in the index's components list on a given date. The data regarding historical components of the index have been sourced from a GitHub repository which included a file with all of the changes in S\&P500 constituents (Farrell 2024). Due to the data availability issues, the stock universe on some dates was smaller than 500, however on each trading day, the majority of stocks have been represented. Adjusted closing prices have been used for both the opening and closing of all of the positions. 
\subsection{Performance metrics}  
\par Comprehensive list of performance metrics will be considered, in order to evaluate the strategy in detail. Aside from metrics used by (Carea et al. 2023), metrics used by (Bieganowski and Ślepaczuk 2024) will also be utilized.
\begin{itemize}
    \item{\textbf{Annualized Return Compounded (ARC)}}
    \par Annualized returns could be the simplest metric informing about the profitability of the strategy. Since in this study, all the returns are reinvested, the compounded version will be considered and used in the calculations of all the metrics involving annualized returns. 
        \begin{equation}
            ARC = \sqrt[(t_n - t_o)]{\frac{V_{t_n}}{V_{t_o}}} - 1
        \end{equation}

        Where:
        \begin{itemize}
            \item{ $(t_n - t_o)$ - Duration of the investment in years}
            \item{ $V_{t_n}$ - Starting capital}
            \item{ $V_{t_o}$ - Final capital}
        \end{itemize}
    \item{\textbf{Annualized Standard Deviation (ASD)}}
    \par Volatility is a standard measure of Portfolio performance. Generally, the higher volatility of a given asset corresponds to a higher associated risk of investing in this asset. Annualized standard deviation is a standard metric of volatility, also used in other metrics. In this paper, it is computed as the standard deviation of daily returns, scaled by a square root of trading days of S\&P500 stocks in a year (252)
    \begin{equation}
        ASD = \sqrt{\frac{\sum_{t=1}^{N}(R_t-\hat{R})^2 }{(N-1)} \cdot n_{days}}
    \end{equation}
    where:
    \begin{itemize}
            \item{ $R_t$ - Return on day t}
            \item{ $\hat{R}$ - Average of daily returns}
            \item{ $n_{days}$ - number of trading days in a year}
    \end{itemize}
        
    \item{\textbf{Information Ratio* (IR*)}}
    \par Sharpe ratio is the most popular measure of risk-adjusted returns, computed as a quotient of the difference between average portfolio returns and the risk-free rate and average standard deviation of the returns. Since the strategy tested in this paper is market-neutral, a risk-free rate is omitted. Moreover, all of the returns are compounded, thus the formula presented below has been used:
    \begin{equation}
        IR^{*} = \frac{ARC}{ASD}
    \end{equation}
    \item{\textbf{Sortino Ratio (ST)}}
    \par Sortino ratio (Sortino 1994) is a modified Sharpe ratio, which only considers the downside deviation of the returns. It is an arguably significantly better metric, as the high deviation of positive returns is unlikely to be a concern for an investor, while the same can't be said about negative returns. It can be calculated as follows.:
    \begin{equation}
        Sortino ratio = \frac{ARC - Risk Free Rate}{Downisde deviation}
    \end{equation}
    \par In this paper the risk-free rate is assumed to be 0 since the strategy is market-neutral.
    \item{\textbf{Max Drawdown (MDD)}}
    \par Maximum drawdown is the largest observed drop of the portfolio's value, expressed as a percentage of the highest value observed before the drawdown. It is a crucial risk metric for through assessment of an investment strategy, as it conveys information about the robustness to tail events, as well as the extent of potential leverage that can be possibly applied to the strategy with no risk of bankruptcy.
    \begin{equation}
        MDD = \max_{\tau \in [0, T]}(\max_{t \in [0, \tau]} \frac{V_t - V_{\tau}}{V_{t}})*100\%
    \end{equation}
    \item{\textbf{Max Loss Duration (MLD)}}
    \par Maximum Loss Duration informs about the longest period in which a portfolio's value has been lower than any previously observed local maximum. It is expressed in years.
    \begin{equation}
        MLD = \frac{t_j - t_i}{n_{obs}}
    \end{equation}
    where: 
    \begin{itemize}
            \item{ $n_{obs}$ - number of trading days in a year}
            \item{ $t_j>t_i$ }
            \item{ $Val(t_j) > Val(t_i)$ }
            \item  Value of the portfolio had local maximums in days $t_j$ and $t_i$
    \end{itemize}
        
    \item{\textbf{Calmar ratio (CR)}}
    \par Calmar Ratio, proposed by Terry W. Young is another useful metric, allowing for assessment of the strategy performance, by focusing on the tail risk measure such as Maximum Drawdown.
    \begin{equation}
            CR = \frac{ARC}{MDD}
    \end{equation}
    \item{\textbf{Information Ratio** (IR**})}
    \par Modified version of Information Ratio, introduced by Kość et al. (2019) will also be applied, as a synthetic measure, able to capture different dimensions of Portfolio performance. In essence, the measure is a product of the Information Ratio and Calmar Ratio, with respect to the direction of the portfolio's value change.
    \begin{equation}
        IR^{**} = \frac{ARC^2 \cdot sign(ARC) \cdot }{ASD \cdot  MDD}
    \end{equation}
\end{itemize}
\subsection {Framework exploration - problem statement}
    \par As stated in the introduction and literature review, The study is based on the strategy proposed by Cartea et al. (2023). Since the authors of the paper used different sources of data and considered a slightly different stock universe, some discrepancies in the results may be present, even in the case of mirroring implementation. To assess the extent of it, the strategy has been tested on the dataset described in 3.1. Precisely, an own implementation has been developed in Python 3, replicating all of the steps and parameters of the strategy discussed in the Cartea et al. (2023) paper. The version involving the use of  SPONGE$_{sym}$ algorithm for clustering, and the number of eigenvectors corresponding to 90\% of explained variance for optimal number of clusters has been selected, as it displayed the best performance in the paper.  Since no information about the transaction costs has been given in the paper, the results have been generated in two variants: with transaction costs equal to 0 and 0.05\%. Strategies has been evaluated with the help of metrics used in the original study.
    \begin{table}[H]
    \centering
    \caption{Performance comparison of strategy implementations and benchmark}
    \label{table:1}
    \begin{tabular}{c c c c c} 
    \hline
    Metric & Original & Own-no-TC &  Own-0.05\%-TC & SPY\\ [0.5ex] 
    \hline
    AR & 12.20 & 10.24 & 2.44  & 6.48\\ 
    SHARPE & 1.10 &  1.17 & 0.28 & 0.33 \\
    SORTINO & 2.01 & 2.09 & 0.49 & 0.51\\
    \hline
    \end{tabular}
    \subcaption*{\footnotesize{Source: Own Elaboration. Table comparing the performance of the original strategy, own replications, and the benchmark, market performance. \textbf{Original} - Results from author's paper. \textbf{Own-no-TC} - Results of the own implementation without transaction costs. \textbf{Own-0.05\%-TC} - Results of the own implementation with 0.05\% transaction costs. \textbf{SPY} - Performance of SPY ETF. \textbf{AR} stands for annualized returns, \textbf{SHARPE} for Sharpe ratio while \textbf{SORTINO} for Sortino ratio. The strategies testing period lasted from 03.2000 to 12.2022.}}
    \end{table}

    \par As can be seen in Table 1, with the assumption of no transaction costs, the values of Sharpe and Sortino ratios as well as annual returns closely match the results obtained by Cartea et al. (2023). The discrepancies could be caused by differences in the stock universe considered or convention of metrics calculation. However, once the transaction costs have been applied, a significant impact on performance can be observed. Figures 1 and 2, attached below further visualize the situation. Both figures visualize the equity curve, which is a function of the value of a portfolio managed using a given strategy over time.
    \begin{figure}[H]
        \centering
        \caption{Equity curve of own implementation with no transaction costs}
        \label{fig:enter-label}
        \includegraphics[width=1\linewidth]{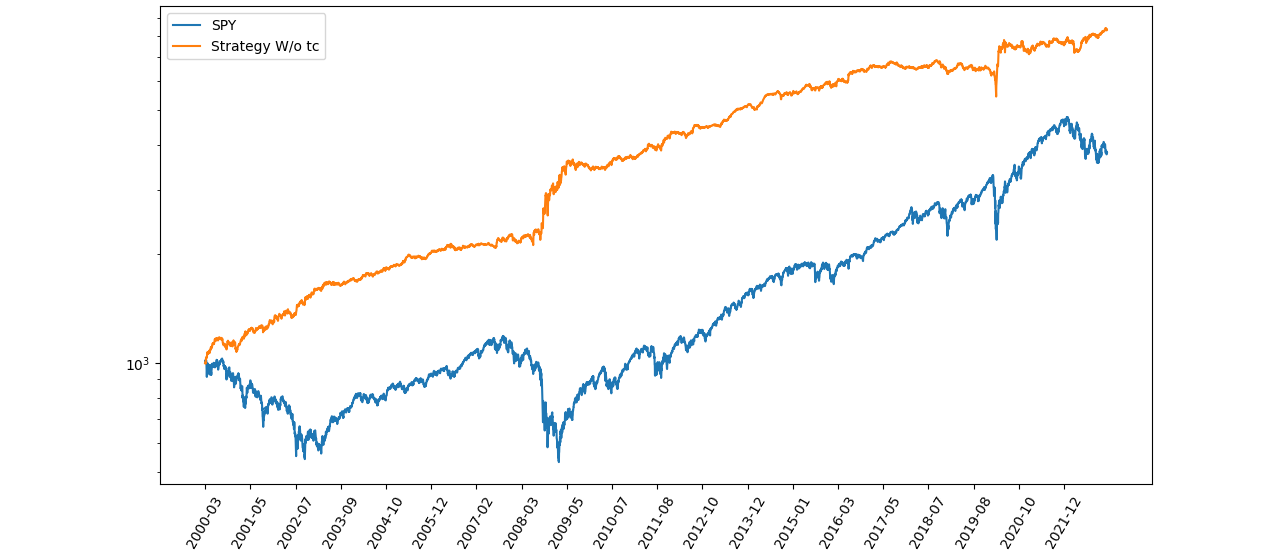}
    \subcaption*{\footnotesize{Source: Own Elaboration. equity curve generated using own backtesting implementation, based on the paper of Cartea et al. (2023)}\textbf{SPY}: SPY ETF, \textbf{Strategy W/o tc}: own implementation of the strategy with no transaction costs}

    \end{figure}
    \begin{figure}[H]
        \centering
        \caption{Equity curve of own implementation with 0.05\% transaction costs}
        \label{fig:enter-label}
        \includegraphics[width=1\linewidth]{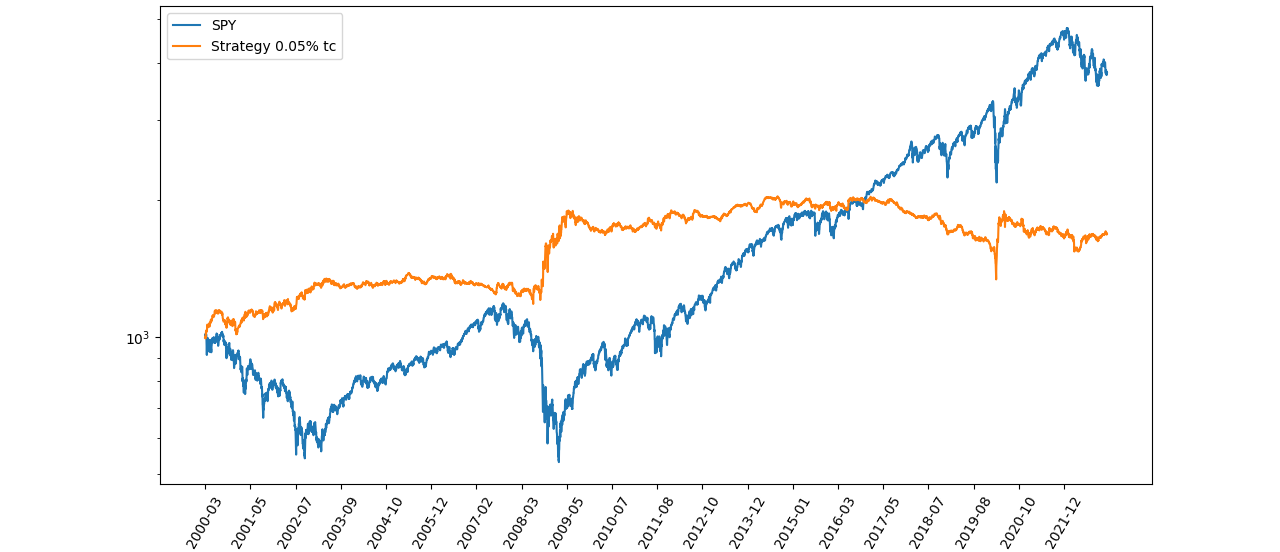}
        \subcaption*{\footnotesize{Source: Own Elaboration. equity curve generated using own backtesting implementation, based on the paper of Cartea et al. (2023). \textbf{SPY}: SPY ETF, \textbf{Strategy 0.05\% tc}: own implementation of the strategy with transaction costs equal to 0.05\%.}}
    \end{figure}
    As can be observed in Figure 1 and Figure 2, the strategy is very susceptible to transaction costs, which effectively turns periods of low growth into periods of losses and stagnation. As can also be observed in Figure 3, The amount of transaction costs incurred is four times larger than the net profit from the strategy, as visible in Figure 2. 
    
    \begin{figure}[H]
        \caption{Curve of transaction costs incurred over time, in the own implementation with 0.05\% transaction costs}
        \label{fig:enter-label}
        \centering
        \includegraphics[width=1\linewidth]{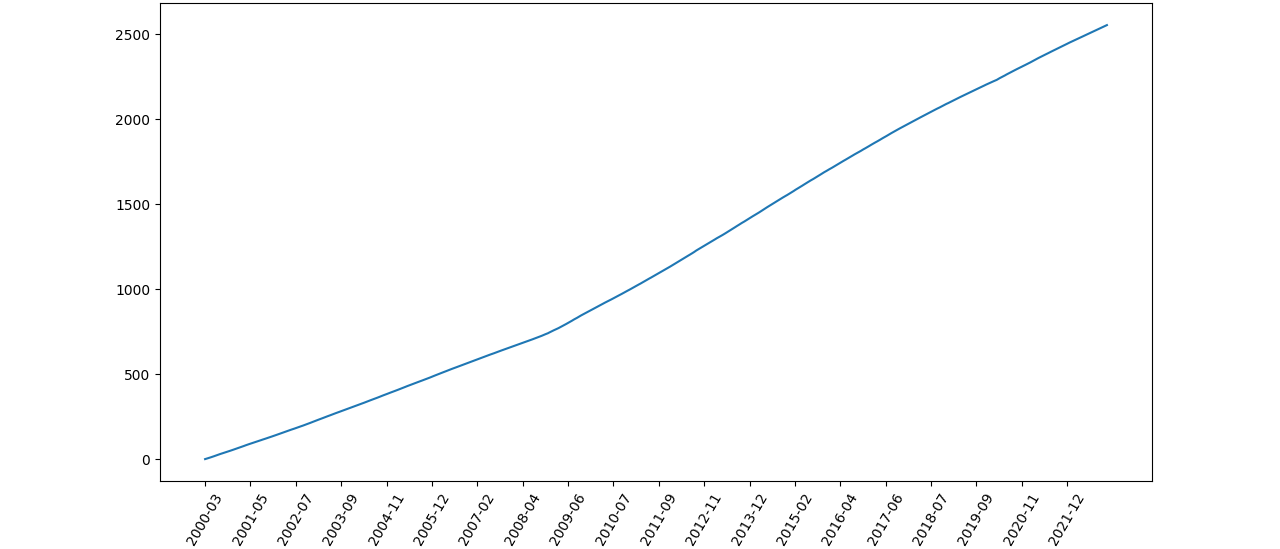}
        \subcaption*{\footnotesize{Source: Own Elaboration. Curve produced using the own backtesting implementation, assuming 0.05\% transaction costs.}}
    \end{figure}
    
    \par However, the transaction costs issue is to a large extent fixable. Fundamentally one needs to reduce the volume traded. Therefore, as a starting point in this research, the described strategy has been altered, such that the rebalance of the portfolio occurs every 10th trading day, instead of every 3rd one. Subsequently, the number of days used in the lookback window to identify clusters of correlated groups of stocks has been also increased (to 30 days, from an initial value of 5).
    \par The more thorough solution to the problem has been proposed after analyzing it from first principles. While extending the time between portfolio rebalances reduces the amount of transaction costs incurred, by limiting the amount of transactions, it needs to be acknowledged that the mere value of costs incurred has secondary importance. The significant decline in the performance of the strategy can to a large extent be attributed to the period in which the transaction costs outweighed gains from the strategy, creating periods of downward trend, not present without transaction costs, but spanning multiple years when transaction costs have been considered. Such periods worsen the performance the most, by impairing annualized returns, increasing downside risk and draw-down, as returns higher than $x\%$ are needed to recover from $x\%$ loss. A potential solution to this issue could be based on improving the average quality of signal, as marginal improvements in that domain could be sufficient to eliminate long periods of downward trend. Therefore, in this study machine learning classifiers will be used to identify signals which are likely to be profitable, and then positions will be opened only for those with a high enough degree of certainty.

\subsection{Features Extraction and Engineering}
\par In order to implement the proposed solution, a specific dataset was needed in order to train classifiers. In order to ensure that both training and testing datasets were representative of various market conditions, the first 1500 trading days starting at the beginning of the year 2000 had been selected as the train data, with the remaining part of the data serving as an out-of-sample test. In order to successfully train classifiers to recognize mean reverting signals, features based on both graph properties and standard price behavior have been extracted. Such an approach was implemented with the goal of capturing both within-cluster relationships and idiosyncratic stocks' behaviors while training the model on a simple format of data, where each signal corresponds to one observation. Below is the list of all features extracted. In line with the interpretation of the clustering algorithm, the naming convention is applied, such that a set of actively traded stocks (constituents of S\&P500 in a given day) is referred to as a graph and the identified cluster is referred to as a sub-graph.
\subsubsection{Graph-based features}
    \begin{itemize}
            \item Local vertex degree:
                \begin{equation}
                \frac{\sum_{n=1}^{S}e_{i,n}-1}{S-1}
                \end{equation}
                Where:
                \begin{itemize}
                \item $e_{i,n}$ = weight of an $n$-th edge coming from an $i$-th vertex
                \item $S$ = sub-graph size
                \end{itemize}
            \item Global vertex degree: 
            \begin{equation}
                \frac{\sum_{n=1}^{G}e_{i,n}-1}{G-1}
            \end{equation}
            Where: 
            \begin{itemize}
                \item $e_{i,n}$ = weight of an $n$-th edge coming from an $i$-th vertex
                \item $G$ = Graph size
            \end{itemize}
            \item Graph density: 
            \begin{equation}
                \frac{\sum_{i=1}^{S}(\sum_{n=1}^{S}{e_{i,n}} - 1)}{(S-1)S}
            \end{equation}
            Where:
            \begin{itemize}
                \item $e_{i,n}$ = weight of an $n$-th edge coming from an $i$-th vertex
                \item $S$ = sub-graph size
            \end{itemize}
            \item Cluster size: 
            \begin{equation}
                \frac{|v_j|}{G}
            \end{equation}
            Where:
            \begin{itemize}
                \item $|v_j|$ = Number of vertices in a $j$-th Cluster
                \item $G$ = Graph size
            \end{itemize}
            \item Number of clusters: \\Number of clusters (sub-Graphs) divided by Graph size 
        \end{itemize}
    As can be observed all of the graph-based features have been normalized by dividing by the term containing either the number of stocks traded (graph size) or by cluster size (sub-graph size). While dividing by cluster size is intended to make features comparable between clusters, dividing by graph size is intended to accommodate the issue of stock universe size being not equal to 500 at times (as mentioned in section 3.1) 
    \subsubsection{Conventional features}
        \begin{itemize}
            \item Cumulative returns deviation from the cluster returns for the last five days
            \item Sign of the deviation from cluster's mean (type of position - long/short)
            \item Mean cluster returns for the last 10 days
            \item Mean stock returns in the last 10 days
        \end{itemize}
\par The dataset has been created by backtesting the version of the original strategy with reduced frequency of rebalances as described in part 3.3. Only the first 1500 trading days have been used, making that period an in-sample one, thus it has not been used for the strategy testing in chapters 4 and 5. For every signal, a record has been stored, which included both the described features as well as binary variables indicating whether the signal has been profitable. Two alternative conditions have been specified for noting the signal as profitable. The first condition assumed signal is profitable if its cumulative returns at the end of a trading day since the last portfolio's rebalance has been greater than the specified threshold $T$. The second condition assumes that the signal is profitable if at the time of a rebalance, cumulative returns since the last rebalance have been greater than the transaction costs (meaning no loss has been incurred on the position). 
\par An important decision was selecting an appropriate value of the threshold $T$.  Figure 4 below displays how the percentage of signals classified as profitable and shares of signals in the case of which two conditions were met changed as the value of $T$ changed. 
\begin{figure}[H]
    \centering
    \caption{Share of signals by type, with different take profit threshold}
    \label{fig:enter-label}
    \includegraphics[width=1\linewidth]{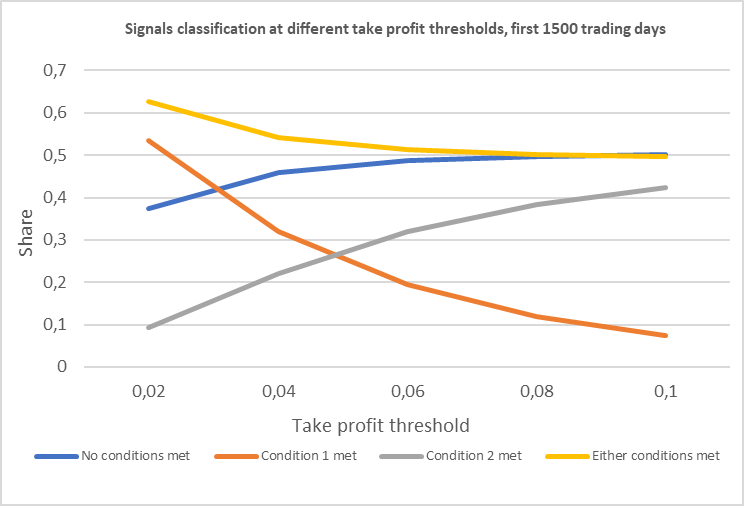}
    \subcaption*{\footnotesize{Source: Own Elaboration. Results generated by backtesting the strategy and noting the percentage of trades meeting given condition depending on the value of the take profit threshold. \textbf{Condition 1 met} - Position closed at take profit. \textbf{Condition 2 met} - Position not closed at take profit but no loss incurred during next rebalance. \textbf{No conditions met} - losses incurred on given signal } \textbf{Either condition met} - Condition 1 or Condition 2 met. results generated based on data from the first 1500 trading days}
\end{figure}
\par As can be observed in Figure 4, with the increase of $T$, the percentage of signals classified as profitable slowly converges to 0.5. What can also be observed is the visible migration of signals from meeting the first condition to meeting the second condition. Overall, since the data was to be used by the machine learning models, the situation with a close to even balance of classes has been identified as a favorable choice. At the same time, choosing too high $T$ could compromise the model's ability to recognize less stable, but profitable signals. Finally, considering such a trade-off, $T$ equal to 4\% has been selected, as it guaranteed the split close to even and represented a significant majority of profitable signals. 
\par After deciding on the value of $T$, the back-testing simulation has been executed, with the parameters discussed in the section 3.3. Thus, a dataset has been constructed in which a single observation consisted of the 14 features listed in the previous paragraph, as well as a binary variable indicating, whether the signal was profitable. This dataset was then scaled using the SKlearn implementation of MinMax Scaler.
\subsection{Ensemble Classifier of signals quality}
\par To maximize the potential of successful classification, five different models have been trained, to identify the best ones and construct an ensemble, to boost the performance of individual models.  In the case of each model, the prepared dataset has been split into training and validation subsets by random sampling, with 80\% and 20\% of samples respectively. Then, in order to optimize performance, grid search cross-validation have been performed, using the standard SKLearn GridSearchCV function. Moreover, Brier score loss has been selected as a metric for hyper parameters tuning, as it optimizes the quality of probability estimates in classification tasks, rather than just the quality of binary class predictions. 
\begin{itemize}
    \item \textbf{Multi-layer Perceptron Classifier}
    \par Multi-layer Perceptron Classifer used, or in short MLP, is a simple, SKLearn native, neural network implementation, which minimizes the logarithmic loss function. It also supports a softmax output layer, thus it's able to return probabilities of belonging to a certain class in classification problems (Pedregosa et al., 2011). The Grid of hyper-parameters searched has been included in Table 2. The set of best parameters, selected through cross-validation, has been bolded.
    \begin{table}[H]
    \centering
    \caption{Hyper-parameters grid- multi-layer perception classifier}
    \label{demo-table}
    \begin{tabular}{ c c c c c c }
    \hline
    Hidden layer sizes & Activation function & Alpha & Learning rate & Batch size & Solver \\
    \hline
    \textbf{64,64} & tanh & 0.1 & \textbf{constant} & 50 & sgd  \\
    64,32 & \textbf{relu}   & 0.01 & adaptive &  100& \textbf{adam}\\
    64 & sigmoid & 0.0001 & &  \textbf{200}   &\\
    32, 16 & &  0.00001 & &   &\\
     &  & \textbf{0.000001}&  & &\\
     &  & 0.0000001&  &  & \\
    \hline
    \end{tabular}
    \subcaption*{\footnotesize{Source: Own Elaboration. hyperparameters used in grid search cross validation on the randomly selected 80\% fraction of the dataset created based on the strategy performance during first 1500 trading days. Chosen hyperparameters has been bolded. }}
    \end{table}
    
    \item \textbf{AdaBoost Classifier}
    \par Ada boost classifier (Freund and Schapire, 1995) is an algorithm in which a sequence of weak learners is fitted, and their predictions are combined in majority voting (Pedregosa et al., 2011). In successive iterations, the weights of samples that were incorrectly classified are being increased, thus forcing the model to learn how to classify difficult cases effectively (Hastie et al., 2009). The Grid of hyper-parameters searched has been included in Table 3. The set of best parameters, selected through cross-validation, has been bolded.
    \begin{table}[H]
    \centering
    \caption{Hyper-parameters grid - AdaBoost Classifier}   
    \label{demo-table}  
    \begin{tabular}{ c c }
    \hline
    number of estimators & learning rate \\
    \hline
    3 & \textbf{0.001}  \\
    5 &  0.01 \\
    10 & 0.1 \\
    20 & 1 \\
    50 & 10 \\
    \textbf{100} &  \\
    \hline
    \end{tabular}
    \subcaption*{\footnotesize{Source: Own Elaboration. hyperparameters used in grid search cross validation on the randomly selected 80\% fraction of the dataset created based on the strategy performance during first 1500 trading days. Chosen hyperparameters has been bolded. }}
    \end{table}
    
    \item \textbf{HistGradientBoosting Classifier}
    \par HistGradientBoosting is another SKlearn implementation of gradient-boosted trees, based on Microsoft's LightGBM. As stated on the SKLearn website, the algorithm is in order of magnitudes faster on datasets containing large numbers of samples. The main difference between the algorithm and other gradient-boosting-based solutions is that it groups data into bins, which limits the number of splits to consider and allows for the use of integer-based data structures rather than continuous ones (Pedregosa et al., 2011).  The Grid of hyper-parameters searched has been included in Table 4. The set of best parameters, selected through cross-validation, has been bolded.
    \begin{table}[H]
    \centering
    \caption{Hyper-parameters Grid, Histogram Gradient Boosting Classifier}
    \label{demo-table}
    \begin{tabular}{ c c c c c c c }
    \hline
    learning rate & early stopping & Max iter &  Warm start \\
    \hline
    0.01 & \textbf{auto }& 50 & \textbf{0}  \\
    \textbf{0.1} & False   & \textbf{100} & 1 \\
    0.5 &  & 150 &  \\
    1 &    & 250 & \\
    5 &   & 500 &\\
    \hline
    \end{tabular}
    \subcaption*{\footnotesize{Source: Own Elaboration. hyperparameters used in grid search cross validation on the randomly selected 80\% fraction of the dataset created based on the strategy performance during first 1500 trading days. Chosen hyperparameters have been bolded. }}
    \end{table}
    \item \textbf{Stochastic Gradient Descent Classifier}
    Although it is not a specific algorithm in itself, Stochastic Gradient Descent is a very efficient approach to fitting linear classifiers, with convex loss functions (Pedregosa et al., 2011). The method has been thus used to add another linear classifier, other than logistic regression.  The Grid of hyper-parameters searched has been included in Table 5. The set of best parameters, selected through cross-validation, has been bolded.
    \begin{table}[H]
    \centering
    \caption{Hyper-parameters grid - Stochastic gradient Descent classifier}
    \label{demo-table}
    \begin{tabular}{c c c c c c c c c}
    \hline
    \makecell{Loss \\ function} & Penalty & Alpha & Max iter & \makecell{Early \\ stopping} & \makecell{Learning \\  rate} & Warm start \\
    \hline
    \textbf{modified Huber} & \textbf{l2} & 0.5 & \textbf{200} & 1 & constant & 1 \\
    log loss & l1   & 0.1 & 1000 & \textbf{ 0} & \textbf{optimal}& \textbf{0}\\
    hinge & elastic net & 0.01 & 10000 &    & adaptive&\\
     & &  \textbf{0.001} & & & invscaling &\\
     &  & 0.0011&  & &&\\
    \hline
    \end{tabular}
    \subcaption*{\footnotesize{Source: Own Elaboration. hyperparameters used in grid search cross-validation on the randomly selected 80\% fraction of the dataset created based on the strategy performance during first 1500 trading days. Chosen hyperparameters have been bolded.}}
    \end{table}

    \item \textbf{Logistic regression}
    \par Logistic regression is a simple linear model used for classification, in the case of which the probabilities of a given observation belonging to a certain class are modeled through the logistic function (Pedregosa et al., 2011).  The Grid of hyper-parameters searched has been included in Table 6. The set of best parameters, selected through cross-validation, has been bolded.
    \begin{table}[H]
    \centering
    \caption{Hyper-parameters grid - Logistics regression}
    \label{demo-table}
    \begin{tabular}{c c c c c c c c c}
    \hline
    C & Penalty & Solver & Max iter & Class weight &  Warm start \\
    \hline
    0.1 & \textbf{l2} & \textbf{lbfgs} & 50 &\textbf{Balanced} & 1 \\
    0.5 & l1   & libliner & \textbf{75} &  None &  \textbf{0}\\
    1 & elastic net & newton-cholesky & 100 &    & \\
    1.5 & &   & 200& & \\
    2 &  & & 500 & &\\
    5 &  & &  & &\\
    \textbf{8} &  & &  & &\\
    10 &  & &  & &\\
    12 &  & &  & &\\
    15 &  & &  & &\\
    20 &  & &  & &\\
    \hline
    \end{tabular}
    \subcaption*{\footnotesize{Source: Own Elaboration. hyperparameters used in grid search cross-validation on the randomly selected 80\% fraction of the dataset created based on the strategy performance during the first 1500 trading days. Chosen hyperparameters have been bolded.}}
    \end{table}
    \end{itemize}

    \textbf{Performance of classifiers}
    \par After hyperparameter tuning and training, the performance of the five models has been examined. Brier Score and precision score have been selected as performance metrics, as they have been identified as most appropriate for the task.
    The results have been presented in Table 7.
    \begin{table}[H]
    \centering
    \caption{Classifiers Performance}
    \label{demo-table}
    \begin{tabular}{c c c}
    \hline
    Classifier & Brier Score & Precision \\
    \hline
    MLP & 0.243 & 0.568 \\
    ADA boost & 0.247  & 0.544\\
    HistGradientBoosting & \textbf{0.218} & \textbf{0.653} \\
    SGD & 0.247 & 0.547\\
    Logistic Regression & 0.249 &  0.580  \\

    \hline
    \end{tabular}
    \subcaption*{\footnotesize{Source: Own Elaboration. Comparison of the performance of individual classifiers on validation dataset, created as remaining 20\% of the generated dataset. The bolded values indicate the best scores}}
    \end{table}
    
    \par As can be seen in Table 7, the Histogram Gradient Boosting algorithm has performed the best in both metrics, significantly better than all the other algorithms. However, since in each case, the value of the Brier score has been lower than 0.25 and the value of precision has been greater than 0.541 (prevalence of positive class in the dataset), all of the classifiers, have outperformed random results, thus learning some patterns. Due to that, an adjusted solution has been proposed. The solution involves the ensemble of all five models, based on the soft voting, in which the Histogram Gradient Boosting classifier would be assigned a higher weight, due to its performance. To set specific weights, simple heuristics of the double weight for the best-performing model have been used. This means, that once all models output the probability that a certain signal is profitable, those values are divided by three in the case of the HistogramGradientBoosting model and by 6 in the case of other models, and then added. However, it is acknowledged that this choice is largely arbitrary, thus, to assess the impact of such a rule, the results for an equally weighted ensemble, ensemble with the triple weight of the best model and the classifier involving just HistogramGradient Boosting have been generated, and compared with the proposed solution in the Sensitivity analysis.
    \par    Another crucial decision that needed to be made in the implementation of the strategy was a selection of the appropriate threshold of probability estimate, above which the signal is considered profitable. The naive way is the selection of the threshold equal to 0.5. However, as pointed out in section 3.3, in order to reduce the impact of transaction costs, classifiers implementation was intended to reduce the number of transactions. As can be seen in Figure 5 below, a naive approach would not limit the amount of transactions to a great extent, as a great majority of signals will still be considered. 
\begin{figure}[H]
        \centering
        \caption{Distribution of predicted probabilities -  Classifiers Ensemble}
        \label{fig:enter-label}
        \includegraphics[width=1\linewidth]{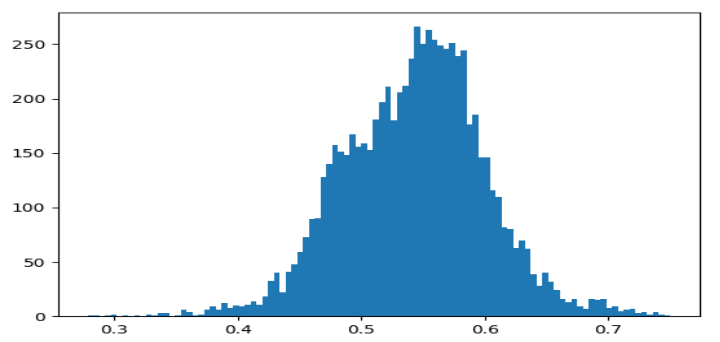}
        \subcaption*{\footnotesize{Source: Own Elaboration. The histogram of probability estimates on validation set returned by soft voting, weighted ensemble of MLPClassifier, AdaBoostClassifier, HistogramGradientBoosting, Logistic Regression, and Stochastic Gradient Descent.HistogramGradientBoosting's weight is doubled in comparison to other weights}}
    \end{figure}
\par Discarding the naive approach, an appropriate value needs to be found for a higher threshold. Sources in literature seem to suggest that generally, a set of 30 to 40 stocks makes a well-diversified portfolio (Statman 1987), thus, selecting a threshold such that on average 10\% of signals will be traded should lead to the construction of a portfolio that will be large enough to fully reap benefits of diversification in a majority of cases.
\par Having decided on the desired number of signals accepted, two possible approaches can be identified. The first approach involves considering certain fixed percentile $P_1$, such that on each portfolio re-balance  $1-P_1$ percent of best-performing stocks is being purchased. The downside of such an approach would be inflexibility, such that numerous positions are still being opened when no very good signals are identified. The second approach involves setting a fixed threshold $P_2$ so that only signals with probability estimates greater than $P_2$ are considered on each rebalance. The inflexibility of the previous one does not burden this approach and, therefore has been selected. In the case of the analyzed ensemble, the 90th percentile of the distribution of predicted probabilities on the validation dataset corresponds to the probability value equal to 0.602, thus, the value of $P_2$ has been rounded up to 0.6. 
\begin{figure}[H]
    \centering
    \caption{Distribution of predicted probabilities - equally weighted ensemble}
    \label{fig:enter-label}
    \includegraphics[width=1\linewidth]{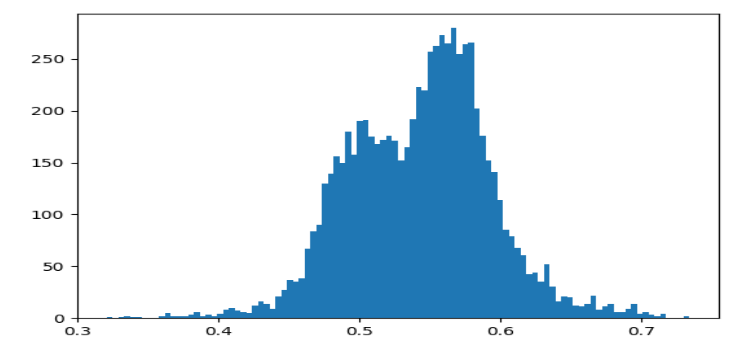}
    \subcaption*{\footnotesize{Source: Own Elaboration. The histogram of probability estimates on validation set returned by soft voting, an equally weighted ensemble of MLPClassifier, AdaBoostClassifier, HistogramGradientBoosting, Logistic Regression, and Stochastic Gradient Descent.}}
\end{figure}
\par The same analysis has been performed for equally weighted Ensemble Figure 6, resulting in finding the 90th percentile of the distribution at 60.08\%, which has also been rounded to 0.6.
\begin{figure}[H]
    \centering
    \caption{Distribution of predicted probabilities - Histogram Gradient Boosting}
    \label{fig:enter-label}
    \includegraphics[width=1\linewidth]{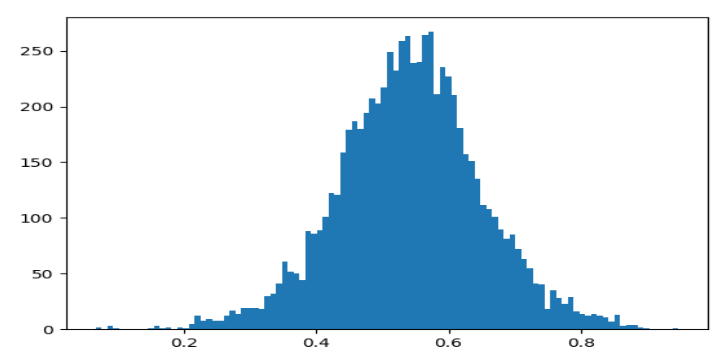}
    \subcaption*{\footnotesize{Source: Own Elaboration. The histogram of probability estimates on the validation set returned by HistogramGradientBoosting classifier}}
\end{figure}
\par In the case of Histogram Gradient Boosting alone, the distribution of predicted probabilities has been more skewed towards the higher values Figure 7. Therefore, applying the same analysis, a 90th percentile equal to 0.679 and the corresponding value of $P_2$ of 0.68 have been selected.

\subsection{implementation and set up}
\par All of the tests have been conducted using the own backtesting implementation. In each testing instance, the equity curve is constructed by starting with 1000 dollars and reinvesting the profits. Moreover, the assumption of the possibility of buying fractional shares is made, as well as the assumption of no excess capital being needed for short positions deposit, thus allowing for zero cost arbitrage at inception. Unless stated otherwise, transaction costs equal to 0.05\% have been applied, both at the closing and opening of the position.

\subsection{Risk management optimizations}

    \subsubsection{Time variant take profit functions}
     \par As discussed earlier the classification of signal quality has been based on the detection of achieving a fixed profit target threshold. An experimental approach to risk management has been tested in the study. The approach aims at hedging against the possibility of signal misclassification, by lowering the threshold with each day passing from opening the position. The take profit function can be expressed with the equation 5:
     \begin{equation}
        Threshold_{tp} = THR * \frac{10-TD}{10} 
     \end{equation}
    where:
    \begin{itemize}
        \item{$TD$ = number of days since the last portfolio rebalance}
        \item{$THR$ = scaling factor}
    \end{itemize}
    To assure congruence with the classifiers, $THR$ has been set to 8\%, as with such value, the expected value of $Threshold_{tp}$ over the 9 days during which the take profit function can be triggered is equal to the threshold used in model's training (4\%). The impact of changing this value has been checked in the sensitivity analysis. The take profit function is only considering the adjusted close prices, on the dates without portfolio rebalance. It means, that the realized profit may be higher, than the  $Threshold_{tp}$ 
\subsubsection{Stop loss functions}
     \par Another risk management optimization applied is the use of stop loss. Although the strategy aims to capitalize on mean reversion effects, thus it seems likely for a position to note increasing loss before the reversal, stop loss could be a helpful tool in preventing abnormal losses during times of market downturn. Similarly to take profit, a time-variant scheme has been tested, with an arbitrary scaling factor equal to 5\%, however, different values have also been inspected in the sensitivity analysis. As in the case of take profit, stop loss is activated based solely on the values of adjusted close prices. Equation 6 represents the Threshold of the applied stop loss function. 
     \begin{equation}
        Threshold_{sl} = 0.05 * \frac{10-TD}{10} 
     \end{equation}
    where:
    \begin{itemize}
        \item{$TD$ = number of days since the last portfolio rebalance}
    \end{itemize}

\subsection{Likelihood weighted decisions}
\subsubsection{Kelly criterion weighted positions}
    \par As stated in the literature review, the Kelly criterion is a formula for optimal sizing of the positions, applicable when one knows the level of risk and wants to maximize the logarithm of wealth in the long term, given that returns are reinvested. In this study, given the overall complexity of the strategy, a simple version of the formula has been applied to each stock separately, to obtain the value of the fraction. Consequently, all of those fractions have been scaled, so that the sums of both shorts' and longs' signals fractions sum up to one. Such scaled fractions represent the share of the current equity assigned to given stocks. Below will be presented the derivation of the optimal fraction value, based on the one presented by (Hagman 2014). 
    A given signal is projected to be profitable with probability $P$. 
    \par Let's assume, that the strategy starts with the initial capital equal to $W_0$. then, assuming the extent of wins and losses is symmetric and given by factor $\lambda$, by investing a fraction $f$, in the $n_{th}$ step, the capital is equal to:
        \begin{equation}
                    W_n = W_0 *(1-\lambda\cdot{f})^{L}(1+\lambda\cdot{f})^{W}
        \end{equation}
    Where L represents the number of losses and W the number of wins. The rate of growth of capital can be given by:
    \begin{equation}
              \frac{W_n}{W_0} = (1-\lambda\cdot{f})^{L}(1+\lambda\cdot{f})^{W}
    \end{equation}

    Average returns, calculated by geometric mean are thus, equal to:
        \begin{equation}
    R = (1-\lambda\cdot{f})^{\frac{L}{n}}(1+\lambda\cdot{f})^{\frac{W}{n}}
        \end{equation}
    The goal is to maximize the value of R. The logarithmic transformation can be applied, as the logarithm is an increasing function. This leads to:
    \begin{equation}
    \log(R) = \frac{L}{n}\cdot \log( 1-\lambda\cdot{f}) + \frac{W}{n}\cdot \log( 1+\lambda\cdot{f})
    \end{equation}
    At this stage, in line with the frequentist approach to probability, it is assumed that the likelihood $P$ of a given signal being profitable, estimated by the classifiers module corresponds to the fraction $\frac{W}{n}$. Consequently, $\frac{W}{n} = P$ and $\frac{L}{n} = 1 - P$. This gives: 
    \begin{equation}
    log(R) = (1-P)\cdot \log( 1-\lambda\cdot{f}) + P\cdot \log( 1+\lambda\cdot{f})
    \end{equation}
    In order to find the value of $f$ which maximizes $R$, the first derivative of $R$ with respect to $f$ is computed, which results in:
    \begin{equation}
    \frac{dR}{df} = \frac{P}{1+\lambda\cdot{f}} - \frac{1-P}{1-\lambda\cdot{f}} = \frac{P - \lambda\cdot{f}\cdot{P}-1+P-\lambda\cdot{f} + \lambda\cdot{f}\cdot{P}}{1 - (\lambda\cdot{f})^2} = \frac{2P -1 - \lambda\cdot{f}}{1 - (\lambda\cdot{f})^2}
    \end{equation}
    Setting the expression to $0$, one obtains:
    \begin{equation}
    f=  \frac{2P-1}{\lambda}
    \end{equation}
    It is easy to check, that the solution is a maximum of $R$.
    \begin{equation}
    \frac{d^2R}{df^2} = \frac{-(1+\lambda\cdot{f})^2-(2P-1 - \lambda)\cdot{f}\cdot{2(1+\lambda\cdot{f}})}{(1-\lambda\cdot{f}^2)^2} <0
    \end{equation}
    Also, since the fractions for all signals are scaled, to sum up to 1, under the assumption of equal $\lambda$, the formula for optimal fraction can be further simplified to:
    \begin{equation}
    f = 2P -1
    \end{equation}
    This result will be applied during the testing of the strategies, and compared with the variant in which all opened positions are equally weighted.
\subsubsection{Risk weighted stop functions}
    \par Another usage of probabilities outputted by the classifier is weighting the take profit and stop loss function. The rationale behind such an approach being similar as with the usage of the Kelly criterion. Through multiplying the take profit function by the likelihood that a signal is profitable, the threshold at which the position is closed is lower for more risky signals, thus the risk of signals closing in loss, due to weak mean reversion is mitigated. Both of the measures described have been tested with the goal of not losing the benefits of diversification while selecting strong signals. The idea has been captured in the equation 16.
    \begin{equation}
        Threshold_{RW} = Threshold \cdot \mathbf{P}
    \end{equation}
    Where:
    \begin{itemize}
     
        \item $Threshold_{RW}$ - Risk-weighted threshold of take profit/stop loss function
        \item $Threshold$ - Threshold of stop loss/take profit function before this modification
        \item $\mathbf{P}$ - Probability that given signal belongs to the class of profitable ones, outputted by the machine learning classifier.
        
    \end{itemize}
\section{Empirical results}
\subsection{Performance of Portfolios} 
\par In this subsection the performance of the the proposed solution is compared with relevant benchmarks. Namely, the strategy is compared with replications of the one proposed by Cartea et al (2023) described in section 3.3, with and without transaction costs. Moreover, SPY ETF has been included in the comparison, serving as a proxy for overall market performance. The table with the computed values of the metrics described in the previous section and the plot of equity curves has been included below.
    \begin{table}[H]
    \centering
    \caption{Performance metrics of the proposed strategy and the relevant benchmarks}
    \label{table:1}
    \begin{tabular}{c c c c c} 
    \hline
    Metric & Strategy & Benchmark & Benchmark W/o tc & SPY \\ [0.5ex] 
    \hline
    ARC & \textbf{49.33\%} & 1.13\% & 9.01\% & 9.12\% \\ 
    ASD & 38.01\% & 9.16\% &  \textbf{9.14\%} & 20.11\% \\
    IR* & \textbf{1.30} & 0.14 & 0.96 & 0.45 \\
    SORTINO & \textbf{3.38} & 0.29 & 1.77 & 0.70 \\
    MDD & 31.98\% & 34.30\% & \textbf{20.68\%} &  55.19\% \\ 
    MLD & 2.10 years & 2.59 years &  \textbf{1.54 years} & 4.85 years \\ 
    CR & \textbf{1.54} & 0.04 & 0.44 & 0.17 \\
    IR** & \textbf{2.00} & 0.01 & 0.42 & 0.08 \\[1ex] 
    \hline
    \end{tabular}
    \subcaption*{\footnotesize{Source: Own Elaboration. Own backtesting implementation, out-of-sample performance between 03.2006 and 12.2022.  \textbf{Strategy}: proposed approach, with weighted ensemble classifier, stop loss and take profit function modifications and kelly citerion. \textbf{Benchmark}: Own implementation of the strategy proposed by Cartea et al. (2023), with 0.05\% transaction costs. \textbf{Benchmark W/o tc}: Own implementation of the strategy proposed by Cartea et al. (2023), with no transaction costs. \textbf{SPY}: SPY ETF. Bolded values indicate the best metric of all presented strategies.}}
    \end{table}
\par As can be seen in Table 8, the proposed strategy is characterized by a high rate of returns and volatility. However, when compared to the benchmarks, the increase in the rate of annualized returns compounded (ARC) is disproportionately higher than the increase in volatility (ASD), resulting in IR* almost ten times higher than that of the benchmark strategy at the same level of transaction costs. Moreover, the strategy's edge over compared approaches is even greater in the case of the Sortino ratio, indicating that the downside volatility is less than overall volatility. Such a situation is favorable for an investor. These results have been achieved with a simultaneous decrease of drawdown when compared to the benchmark and the overall market (represented by SPY). Therefore, the proposed strategy is characterized by visibly higher values of calmar ratio and Information Ratio**. 
\par To ensure the significance of obtained results, a probabilistic t-test has been conducted, to compare information ratios of the proposed strategy and the one of the SPY ETF. The critical value has been set at 0.01. The test has been constructed like the one applied by Bieganowski \& Ślepaczuk (2024), meaning that, it was calculated by the formula: 
\begin{equation}
    t= \frac{IR^*_{s}-IR^*_{b}}{\frac{\sigma}{\sqrt{n}}}
\end{equation}
Where:
\begin{itemize}
    \item $n$ - sample size
    \item $\sigma $ represents the standard deviation of the difference between the benchmark and the strategy returns
    \item $IR^*_{s}$ - represents IR* of the proposed strategy
    \item $IR^*_{b}$ - represents IR* of benchmark
\end{itemize} 
\par     The p-values were then calculated based on the Student's t-distribution, with 4225 degrees of freedom.
\begin{table}[H]
    \centering
    \caption{T-test comparing information ratios of the proposed  strategy and S\&P500 benchmark}
    \label{tab:my_label}
    \begin{tabular}{c c c c c c}
            \hline
         IR*-Strategy & IR*-SPY & SE & df & t & p-value  \\
         \hline
         1.30 & 0.45 &  0.0004 & 4225 &  2118.84 &  $< 0.00001$   \\
         \hline
    \end{tabular}
    \subcaption*{\footnotesize{Source: Own Elaboration.   \textbf{IR*-Strategy}: IR* of the proposed strategy. \textbf{IR*-SPY}: IR* of the SPY ETF. \textbf{SE} - standard error. \textbf{df} - number of the degrees of freedom. t - the value of the t-test statistics. \textbf{$H_0$} - IR* of the proposed strategy is not greater than that of SPY ETF}}
\end{table}
\par The null hypothesis stated that the Information Ratio* of the proposed strategy is not greater than that of the SPY benchmark. Results obtained in Table 9 indicate that the null hypothesis can be rejected at a 1\% significance level indicating the strong significance of obtained results.

    \begin{figure}[H]
        \centering
        \caption{equity curves of the proposed strategy and the relevant benchmarks}
        \label{fig:enter-label}
        \includegraphics[width=1\linewidth]{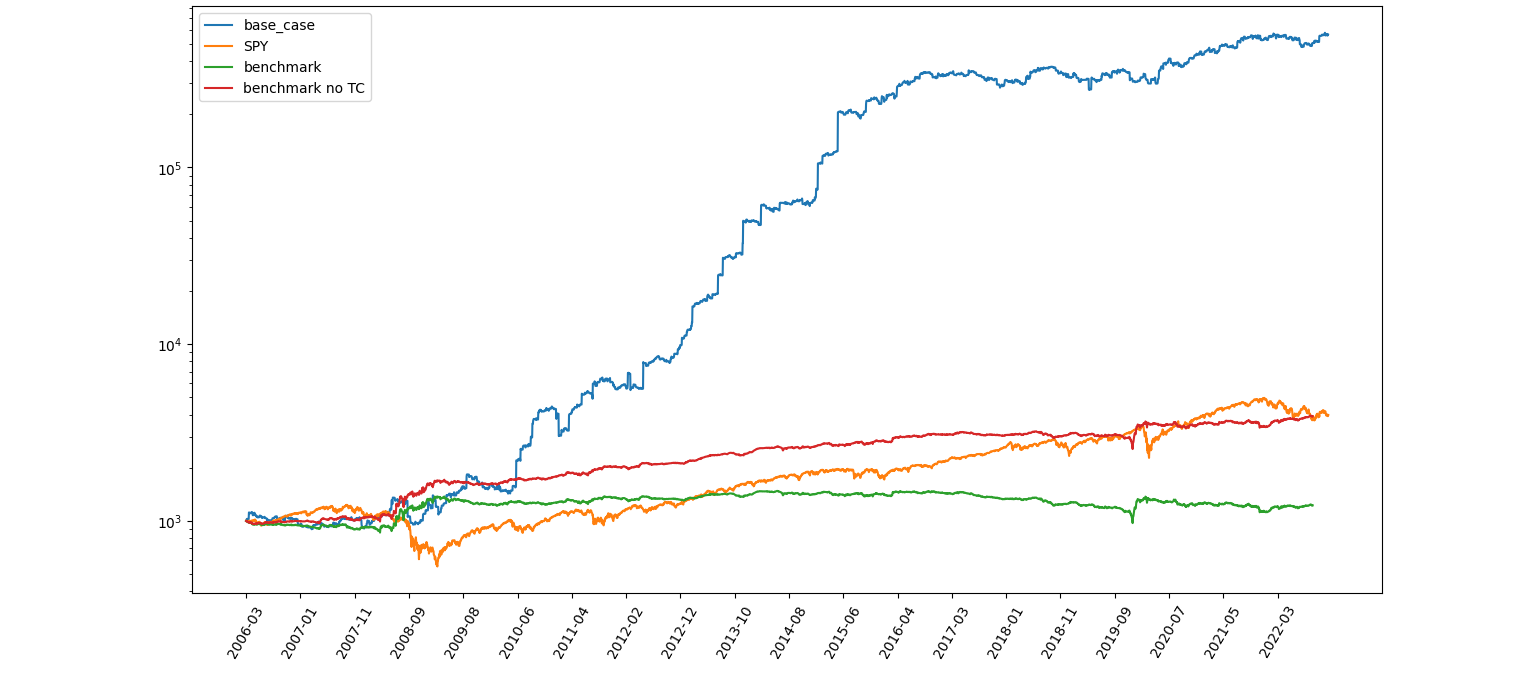}
        \subcaption*{\footnotesize{Source: Own Elaboration. Own backtesting implementation, out-of-sample performance between 03.2006 and 12.2022}}
    \end{figure}
As can be seen in Figure 8, the proposed strategy generates substantially higher returns, though the rate of growth slowed down around the year 2016. An analysis of this phenomenon led to the identification of two factors. Firstly, the starting-point strategy of Cartea et al (2023) also notices a visible decline in performance in a similar time frame. Secondly, a very high rate of returns in the 2010-2016 period can be to a significant extent attributed to the limited number of very profitable transactions. This behavior of the model also leads to a largely 'step-like' structure of the equity curve. Through monitoring of the transactions closed after fulfilling the take-profit criterion, it has been discovered that the stock 'Emporium Petroleum' with ticker EP has been responsible for the majority of those occurrences. Although not frequently, the situation also occurred in the case of the stock of the company: 'Ocean Thermal Energy Corp' (ticker CPWR). Although the companies have indeed been a part of S\&P500 during the testing period, their stocks have been characterized by high volatility with very low to moderate volume. Therefore, even though earning on significant price movements is desirable behavior, the situation in which such a situation is the main driver of the performance is not desirable, thus the performance of the strategy on the datasets with EP and CPWR stocks excluded has been tested in the sensitivity analysis. On the other hand, it could mean that the proposed strategy has a very big potential of gaining a lot on stocks with high volatility and should encourage the potential readers to test it on stocks with lower capitalization and higher volatility than the ones included in the S\&P500 index, eg from RUSSEL2000 or similar indices.
\section{Sensitivity analysis}
    \subsection{Stop functions modifications and the Kelly criterion}
    \par In this subsection performance of the proposed strategy (referred to as the base case from this point) is compared with the performance of strategies that do not use the modifications described in parts 3.6 and 3.7. The first one is the base case strategy without time-variant and probability-weighted modifications in their stop-loss and take-profit functions, referred to as a flat base case with Kelly. The second one, on top of the changes implemented in the first one, doesn't utilize the Kelly criterion and thus is referred to as a flat base case. 
    \begin{table}[H]
    \centering
    \caption{Performance metrics for sensitivity analysis of the stop loss and take profit functions modifications and the Kelly criterion}
    \label{table:1}
    \begin{tabular}{c c c c c} 
    \hline
    Metric & Base case & Flat Base Case Kelly & Flat Base Case & SPY \\ [0.5ex] 
    \hline
    ARC & 49.33\% & \textbf{52.63\%} & 50.33\% & 9.12\% \\ 
    ASD & 38.01\$ & 40.29\% &  38.95\% & \textbf{20.11\%} \\
    IR* & 1.30 & \textbf{1.31} & 1.29 & 0.45 \\
    SORTINO & \textbf{3.38} & 3.25 & 3.26 & 0.70 \\
    MDD & \textbf{31.98\%} & 36.05\% & 33.94\% &  55.19\% \\ 
    MLD & 2.10 years & 2.82 years &  \textbf{1.99 years} & 4.85 years \\ 
    CR & \textbf{1.54} & 1.46 & 1.48 & 0.17 \\
    IR** & \textbf{2.00} & 1.91 & 1.90 & 0.08 \\[1ex] 
    \hline
    \end{tabular}
    \subcaption*{\footnotesize{Source: Own Elaboration. Own backtesting implementation, out-of-sample performance between 03.2006 and 12.2022. \textbf{Base Case}: proposed approach, with weighted ensemble classifier, stop loss and take profit function modifications and Kelly criterion. \textbf{Flat Base case Kelly}: Base case strategy without time variant stop function modifications. \textbf{Flat Base case}: Base Case strategy without time variant take profit and stop loss functions and Kelly criterion. \textbf{SPY}: SPY ETF. Bolded values indicate the best metric of all presented strategies.}}
    \end{table}
    \par As can be seen in Table 10, the performance of the base case and its two modifications are very similar, as the values of IR* differ by not more than 2\% between them. The results suggest that the addition of the Kelly criterion increased the rate of returns at the expense of higher volatility and longer maximum drawdown duration. At the same time, modifying the stop loss and take profit functions reduced volatility and returns by roughly the same factor, but also led to lower values of maximum drawdown, and downside risk allowing the base case strategy to achieve slightly higher values of Calmar, Sortino, and IR** ratios. Reduction of the tail risk is desirable, implying the validity of the time-variant and probability-weighted approach, however, the differences are marginal. Thus it can be said that the strategy proposed is not sensitive to the removal of these modifications.
    \begin{figure}[H]
        \centering
        \caption{equity curves for sensitivity analysis of the stop loss and take profit functions modifications and the Kelly criterion}
        \label{fig:enter-label}
\includegraphics[width=1\linewidth]{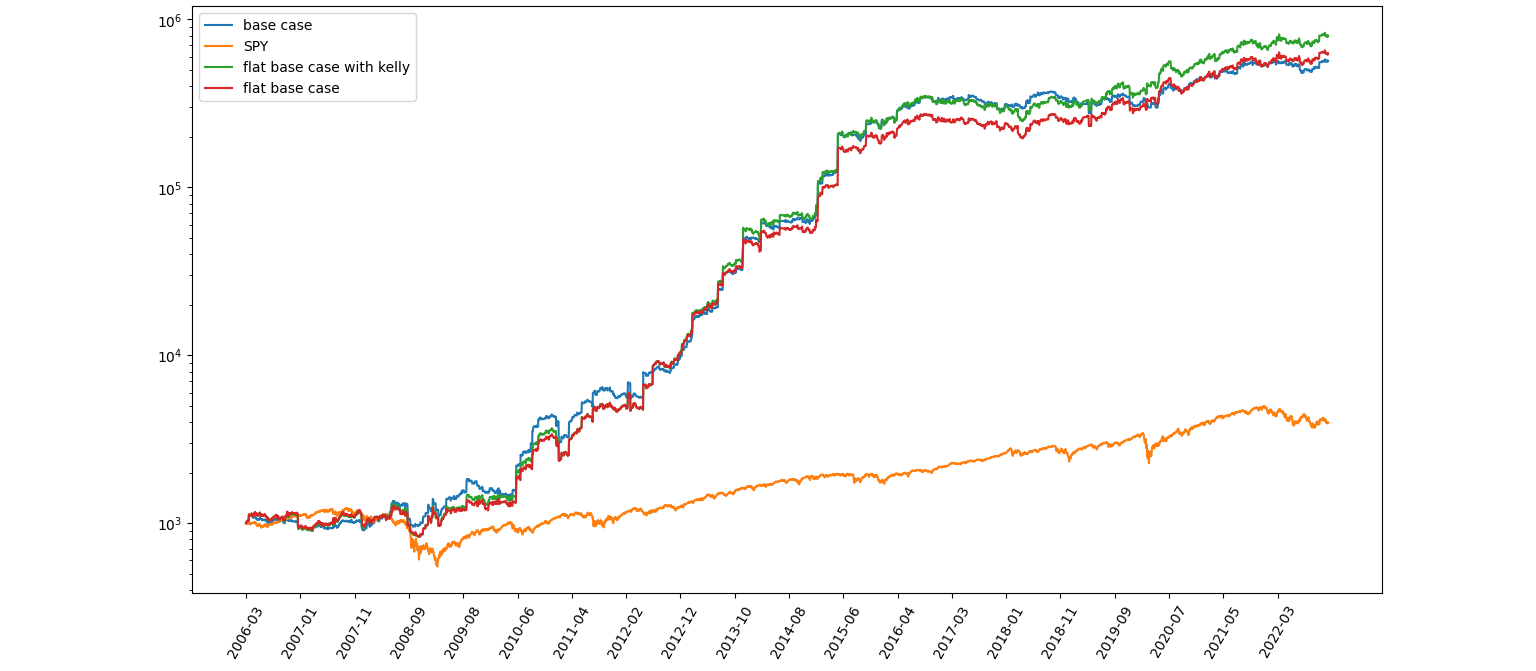}
        \subcaption*{\footnotesize{Source: Own Elaboration. Own backtesting implementation, out-of-sample performance between 03.2006 and 12.2022}}
    \end{figure}
    \par As can be seen in Figure 9, the equity curves of base case strategy and flat versions follow very similar trajectories.
    \subsection{Ensemble Construction}
     \par The results presented in Table 11, clearly suggest that the ensemble weights selected for the base case strategy were the optimal ones. The results obtained with modified weights are almost uniformly worse on every metric than the ones of the base case. At the same time, both alternative strategies proposed are characterized by better values of the performance metrics than the benchmark SPY strategy. As can be further observed in Figure 10, an equally weighted ensemble is characterized by higher volatility than the base case, which is congruent with higher values of ASD and MDD in Table 11. Such results may suggest, that the relative increase of weights of worse performing models leads to lower average quality of predictions, and thus more unprofitable positions have been opened. In contrast, although the equity curve of the strategy in which only Histogram GradientBoosting has been utilized looks relatively stable, it has a significant decrease in performance around the year 2017, which has been identified before as typical for both the base case strategy as well as the strategy presented by Cartea et al (2023). Thus, it can be deduced, that not using a diversified ensemble, when performing the classification of signals quality, may increase the risk of reinforcing some undesirable characteristic of the underlying strategy. Interestingly, the results of the base case strategy but without rounding the threshold used for limiting the number of transactions, differ slightly from the base case, indicating the sensitivity of the strategy to changes in the number of stocks traded.
    \begin{table}[H]
    \centering
    \caption{Performance metrics for sensitivity analysis of the ensemble construction}
    \label{table:1}
    \begin{tabular}{c c c c c c} 
    \hline
    Metric & Base Case & HistGradientBoosting & equally weighted  & W/t rounding & SPY\\ [0.5ex] 
    \hline
    ARC & \textbf{49.33\%} & 31.58\% & 28.91\% & 47.15\% & 9.12\% \\ 
    ASD & 38.01\$ & 34.82\% &  38.22\% & 38.99\% & \textbf{20.11\%} \\
    IR* & \textbf{1.30} & 0.91 & 0.76 & 1.21& 0.45 \\
    SORTINO & \textbf{3.38} & 1.89 & 1.54 & 3.11 & 0.70 \\
    MDD & \textbf{31.98\%} & 50.7\% & 63.95\% & 38.83\%&  55.19\% \\ 
    MLD & \textbf{2.10 years} & 3.01 years &  2.63 years & 3.22 years& 4.85 years \\ 
    CR & \textbf{1.54} & 0.62 & 0.45 & 1.38 & 0.17 \\
    IR** & \textbf{2.00} & 0.56 & 0.34 & 1.67 & 0.08 \\[1ex] 
    \hline
    \end{tabular}
    \subcaption*{\footnotesize{Source: Own Elaboration. Out-of-sample performance between 03.2006 and 12.2022, own backtesting implementation. \textbf{Base case}: proposed approach, with weighted ensemble classifier, stop loss and take profit function modifications and Kelly citerion. \textbf{HistGradientBoosting}: Base case but using only one model rather than an ensemble. \textbf{equally weighted}: The base case with equally weighted ensemble. \textbf{W/t rounding}: Base case but without rounding the threshold used in determining whether a signal is profitable. \textbf{SPY}: SPY ETF. Bolded values indicate the best metric of all presented strategies.}  }
    \end{table}
    \begin{figure}[H]
        \centering
        \caption{equity curves for sensitivity analysis of the ensemble Weights }
        \label{fig:enter-label}
        \includegraphics[width=1\linewidth]{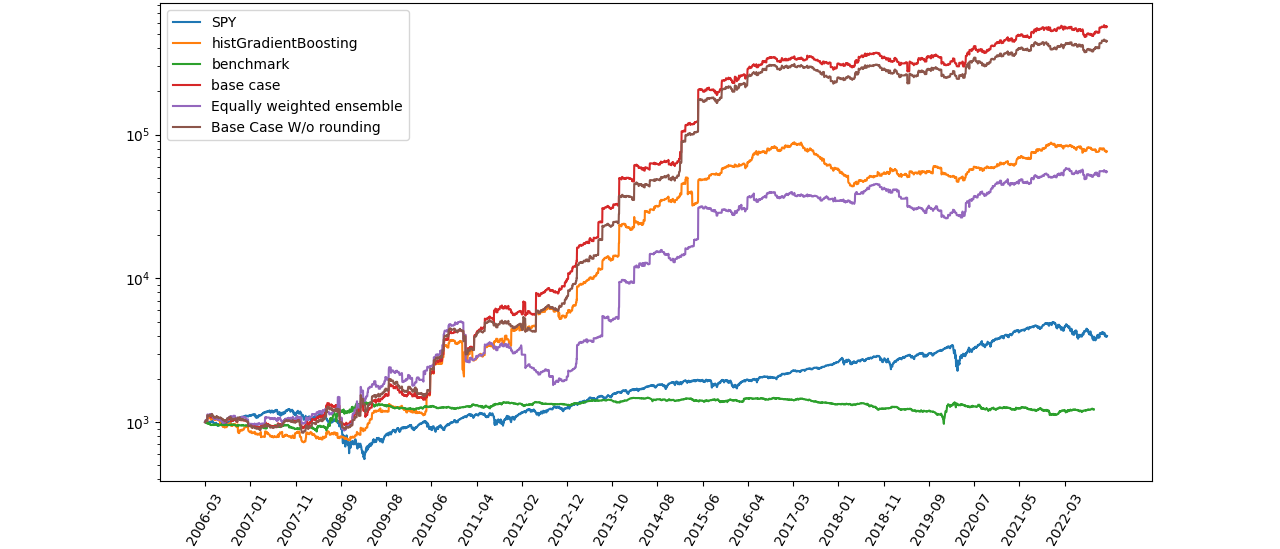}
        \subcaption*{\footnotesize{Source: Own Elaboration. Own backtesting implementation, out-of-sample performance between 03.2006 and 12.2022.}}
    \end{figure}
   
    \subsection{Transaction costs}
    \par The sensitivity analysis of the proposed approach performance at different transaction cost rates, informs us that the base case strategy is largely immune to changes in this parameter. As can be observed in Table 12, as the transaction costs increase the strategy displays monotonically worse performance. At the same time, performance with transaction costs equal to 0.1\% is significantly better than that of SPY ETF. The similarity of the performance is very visible in Figure 11. Thus, the low sensitivity of the base case strategy to transaction costs can be concluded.
    \begin{table}[H]
    \caption{Performance metrics for sensitivity analysis of the transaction cost rate}
    \label{table:1}
    \centering
    \begin{tabular}{c c c c c c } 
    \hline
    Metric & 0\% & Base Case &   0.075\%  & 0.1\% & SPY\\ [0.5ex] 
    \hline
    ARC &\textbf{ 54.55\%} & 49.33\% & 46.78\% & 44.27\% & 9.12\%\\ 
    ASD & 38.02\% & \textbf{38.01\%} & 38.02\% & 38.02\% & 20.11\%\\
    IR* & \textbf{1.43} & 1.30 & 1.23 & 1.16 & 0.45\\
    SORTINO & \textbf{3.74} & 3.38 & 3.20 & 3.027 & 0.70\\
    MDD & \textbf{31.72\%} & 31.98\% & 32.11\% & 32.24\% & 55.19\%\\ 
    MLD & \textbf{1.98 years} & 2.10 years & 2.14 years & 2.17 years & 4.85 years\\ 
    CR & \textbf{1.72} & 1.54 & 1.46 & 1.37 & 0.17 \\
    IR** & \textbf{2.46} & 2.00 & 1.80 & 1.59 & 0.08\\[1ex] 
    \hline
    \end{tabular}
    \subcaption*{\footnotesize{Source: Own Elaboration. Own backtesting implementation, out-of-sample performance between 03.2006 and 12.2022. \textbf{0\%}: Base Case with no transaction costs. \textbf{Base case}: proposed approach, with weighted ensemble classifier, stop loss and take profit function modifications and Kelly criterion.  \textbf{0.075\%}: Base Case with 0.075\% transaction costs. 
    \textbf{0.1\%}: Base Case with 0.1\% transaction costs.\textbf{SPY}: SPY ETF. Bolded values indicate the best metric of all presented strategies.}}
    \end{table}
    \begin{figure}[H]
        \centering
        \caption{equity curves for sensitivity analysis of the transaction cost rate}
        \label{fig:enter-lab}
        \includegraphics[width=1\linewidth]{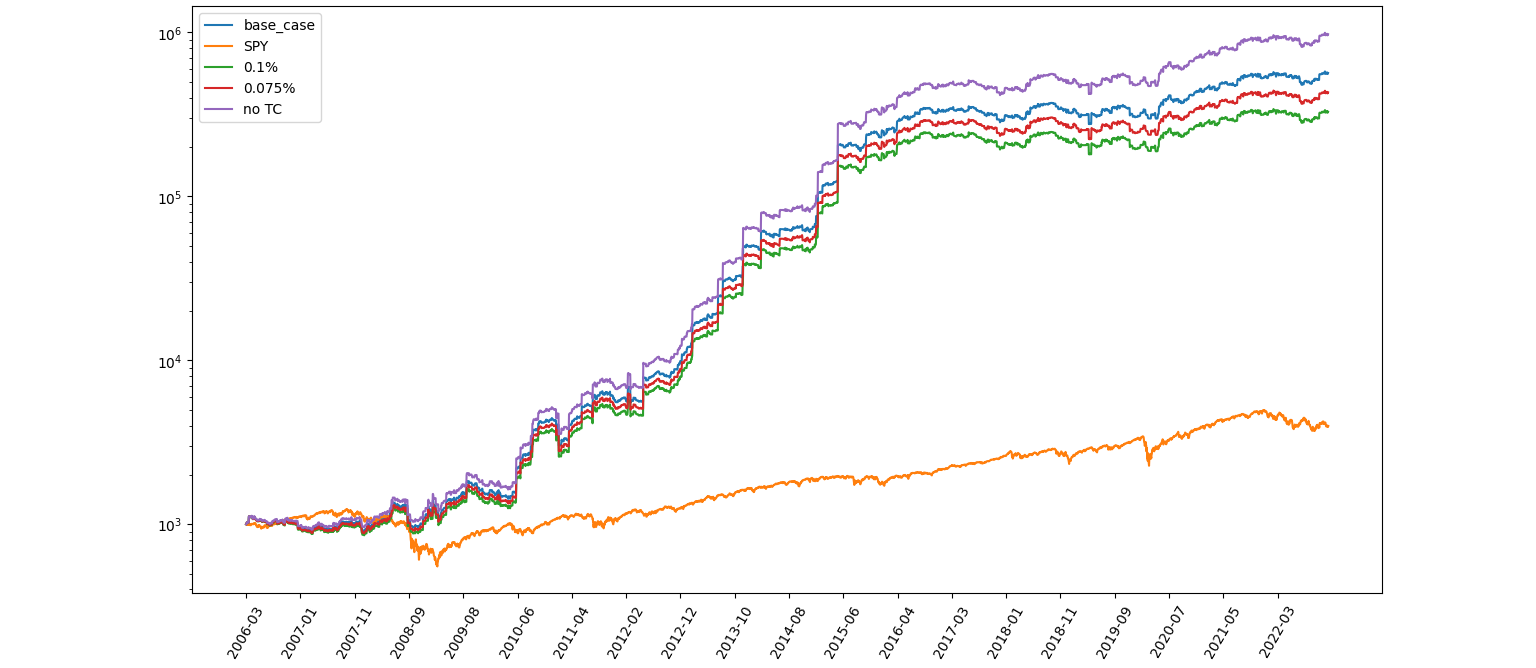}
        \subcaption*{\footnotesize{Source: Own Elaboration. Own backtesting implementation, out-of-sample performance between 03.2006 and 12.2022}}
    \end{figure}
    
\subsection{Stop Loss Threshold}
 \par As can be seen in Table 13, both lowering and increasing the stop loss threshold seem to have a moderate negative impact on the performance of the strategy. While a slight decrease of the threshold to 3\% led to a decrease in maximum drawdown, no other benefits can be observed. Those results align with the expectations, given the strategy design. from one perspective, if the strategy aims to capitalize on mean reversion, having a stop loss threshold that is too low may lead to premature closure of possibly profitable positions. Conversely, a threshold that is too high may simply lead to insufficient reduction of downside risk. However, as previously stated, the strategy is only moderately sensitive to the changes of the threshold.
    \begin{table}[H]
    \centering
    \caption{Performance metrics for sensitivity analysis of stop loss threshold value}
    \label{table:1}
    \begin{tabular}{c c c c c c} 
    \hline
    Metric & 1\% &  3\% & base case & 10\% & SPY\\ [0.5ex] 
    \hline
    ARC & 41.54\% & 41.93\% & \textbf{49.33\%} &  46.00\% & 9.12\%\\ 
    ASD & \textbf{35.29\%} & 35.75\% & 38.01\%  & 39.12\% & 20.11\%\\
    IR* & 1.18 & 1.17 & \textbf{1.30} & 1.18 & 0.45\\
    SORTINO & 3.21 & 3.14 & \textbf{3.38 }& 2.84 & 0.70\\
    MDD & 32.08\% & \textbf{28.58\%} & 31.98\% & 41.05\% & 55.19\%\\ 
    MLD & 3.01 years & 3.75 years & \textbf{2.10 years} & 3.28 years & 4.85 years \\ 
    CR & 1.30 & 1.52 & \textbf{1.54 }& 1.12 & 0.17\\
    IR** & 1.53 & 1.78 & \textbf{2.00} & 1.32 & 0.08\\[1ex] 
    \hline
    \end{tabular}
    \subcaption*{\footnotesize{Source: Own Elaboration. Own backtesting implementation, out-of-sample performance between 03.2006 and 12.2022. \textbf{1\%}: Base Case with stop loss threshold equal to 1\%. \textbf{3\%}: Base Case with stop loss threshold equal to 3\%. \textbf{Strategy}: proposed approach, with weighted ensemble classifier, stop loss and take profit function modifications and Kelly citerion. \textbf{10\%}: Base Case with stop loss threshold equal to 10\%. \textbf{SPY}: SPY ETF. Bolded values indicate the best metric of all presented strategies.}}
    \end{table}
    \begin{figure}[H]
        \centering
        \caption{equity curves for sensitivity analysis of stop loss threshold value }
        \label{fig:enter-label}
        \includegraphics[width=1\linewidth]{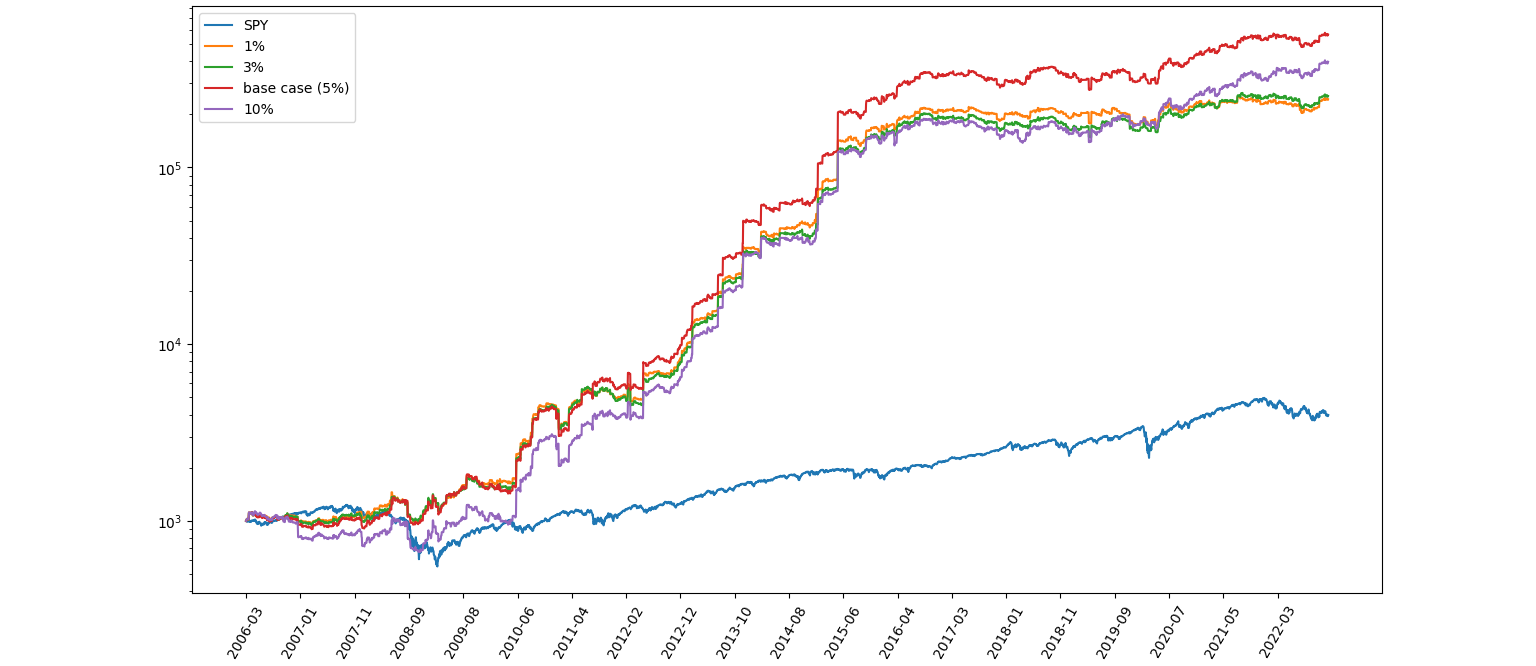}
        \subcaption*{\footnotesize{Source: Own Elaboration. Own backtesting implementation, out-of-sample performance between 03.2006 and 12.2022}}
    \end{figure}
\par As visible in Figure 12, all stop-loss thresholds tested underperform the base case, but still significantly overperform the SPY benchmark.
\subsection{Over-performing stocks exclusion}
 \par As can be observed in Table 14, excluding best-performing stocks has a drastic effect on the performance of the strategy. Given the nature of the stocks excluded, those findings strongly suggest, that the base case strategy's main advantage lays in ability to successfully capitalized on extensive price movements. What needs to be noted, is the fact, that all three variants of the strategy with best-performing stocks excluded, outperform both the benchmark strategy of Cartea et al. (2023) with transaction costs applied and the SPY ETF performance, in terms of the Calmar, Sortino, and information ratios. Those results evince the presence of a performance improvement which can be attributed to the usage of signals quality classifiers. Concurrently, the very high sensitivity of the base case strategy to the exclusion of the highly volatile, profit-driving stocks, highlights both potential issues that can be expected with the scaling of the strategy, as well as the potential opportunities associated with the application of the strategy to more volatile markets.
    \begin{table}[H]
    \centering
    \caption{Performance metrics for sensitivity analysis of over-performing stocks exclusion}
    \label{table:1}
    \begin{tabular}{c c c c c c c} 
    \hline
    Metric & Base case & Benchmark &  W/o  EP \& CPWR &  W/o EP & W/o CPWR &SPY \\ [0.5ex] 
    \hline
    ARC & \textbf{49.33\%} & 1.13\% & 10.73\%  & 17.78\% & 26.65\% & 9.12\% \\ 
    ASD & 38.01\$ & \textbf{9.16\% }&  19.71\% & 22.72\% & 33.38\% & 20.11\% \\
    IR* & \textbf{1.30} & 0.14 & 0.54 &  0.78 & 0.80 & 0.45 \\
    SORTINO & \textbf{3.38} & 0.29 & 0.83 & 2.09 & 1.57& 0.70 \\
    MDD & 31.98\% & 34.30\% & 51.44\% & \textbf{24.95\%} &  50.51\% & 55.19\% \\ 
    MLD & 2.10 years & 2.59 years &  9.38 years& \textbf{1.67 years} & 2.10 years& 4.85 years \\ 
    CR & \textbf{1.54} & 0.04 & 0.21 & 0.71 & 0.53 & 0.17 \\
    IR** & \textbf{2.00 }& 0.01 & 0.11 &0.55 & 0.42 & 0.08 \\[1ex] 
    \hline
    \end{tabular}
    \subcaption*{\footnotesize{Source: Own Elaboration. Own backtesting implementation, out-of-sample performance between 03.2006 and 12.2022. \textbf{Strategy}: proposed approach, with weighted ensemble classifier, stop loss and take profit function modifications and Kelly criterion. \textbf{Benchmark}: Own implementation of the strategy proposed by Cartea et al. (2023), with 0.05\% transaction costs. \textbf{W/o EP \& CPWR}: Base Case with EP and CPWR stocks excluded from the dataset. \textbf{W/o EP }: Base Case with EP stock excluded from the dataset. \textbf{W/o CPWR}: Base Case with CPWR stock excluded from the dataset. \textbf{SPY}: SPY ETF. Bolded values indicate the best metric of all presented strategies.}}
    \end{table}
    \begin{figure}[H]
        \centering
        \caption{equity curves for sensitivity analysis of over-performing stocks exclusion}
        \label{fig:enter-label}
        \includegraphics[width=1\linewidth]{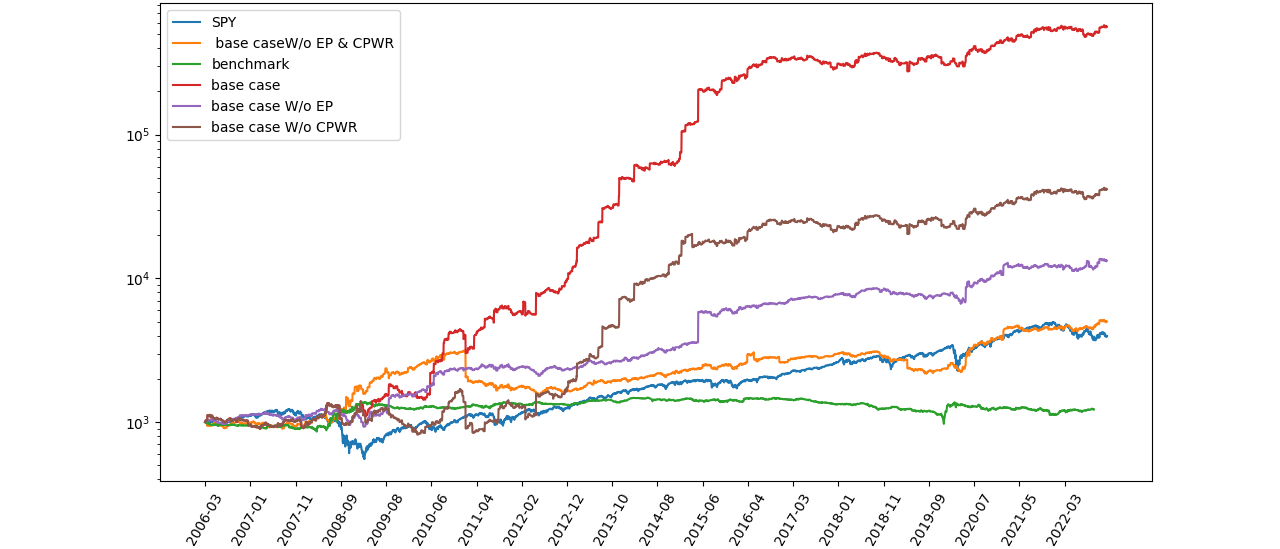}
        \subcaption*{\footnotesize{Source: Own Elaboration. Own backtesting implementation, out-of-sample performance between 03.2006 and 12.2022}}
    \end{figure}
    \par The equity curves in Figure 13 clearly demonstrate the decrease in the performance of the strategies when compared to the base case. However, is the drastic drop of the equity curve of the strategy with both EP and CPWR stocks excluded, around the end of 2010 is worth more detailed consideration. Before this significant drawdown, the strategy has been outperforming all other approaches tested, displaying resilience to the 2008 and 2020 crises. Moreover, it also seems to be less susceptible than the base case to the performance drop after the year 2016. It could be speculated, that with a focus on the parameters tuning for specifically this variant of the strategy, a very stable, and profitable strategy could be achieved. Moreover, this hypothesis can be supported by comparatively low values of maximum drawdown and maximum loss duration, observed in the case of strategy in which only the EP stock has been excluded.
    \subsection{Take profit threshold}
    \par The values of IR* and Sortino for strategies with both halved and doubled take profit threshold when compared to Base case, have changed by less than 10\%. One noticeable change is around 50\% increase of MDD in the case of strategy with a scaling factor equal to 0.16. Overall, as shown in Figure 14, the strategy is not very sensitive to changes in this parameter.
    \begin{table}[H]
    \centering
    \caption{Performance metrics for sensitivity analysis of different take profit scaling factors}
    \label{table:1}
    \begin{tabular}{c c c c c c} 
    \hline
    Metric & Base Case & Benchmark &  0.04 &  0.16 & SPY \\ [0.5ex] 
    \hline
    ARC & 49.33\% & 1.13\% & 45.99\% & \textbf{49.86\%} & 9.12\%  \\ 
    ASD & 38.01\$ & \textbf{9.16\%} & 36.91\% & 39.46\% & 20.11\% \\
    IR* & \textbf{1.30} & 0.14 &  1.24 & 1.26 & 0.45 \\
    SORTINO &\textbf{ 3.38} & 0.29 & 3.32 & 3.10 & 0.70 \\
    MDD & 31.98\% & 34.30\% & \textbf{31.75\% }& 48.43\% &  55.19\% \\ 
    MLD & \textbf{2.10 years} & 2.59 years & 3.72 years & 3.37 years & 4.85 years \\ 
    CR & \textbf{1.54} & 0.04 & 1.45 &1.02 & 0.17 \\
    IR** & \textbf{2.00} & 0.01  & 1.80 & 1.23 & 0.08 \\[1ex] 
    \hline
    \end{tabular}
    \subcaption*{\footnotesize{Source: Own Elaboration. Own backtesting implementation, out-of-sample performance between 03.2006 and 12.2022. \textbf{Base Case}: proposed approach, with weighted ensemble classifier, stop loss and take profit function modifications and Kelly citerion. \textbf{Benchmark}: Own implementation of the strategy proposed by Cartea et al. (2023), with 0.05\% transaction costs. \textbf{0.04}: Base Case with take profit scaling factor equal to 0.04. \textbf{0.16}: Base Case with take profit scaling factor equal to 0.16.\textbf{SPY}: SPY ETF. Bolded values indicate the best metric of all presented strategies.}}
    \end{table}
    \begin{figure}[H]
        \centering
        \caption{Equity curves for sensitivity analysis of the take profit scaling factor}
        \label{fig:enter-label}
        \includegraphics[width=1\linewidth]{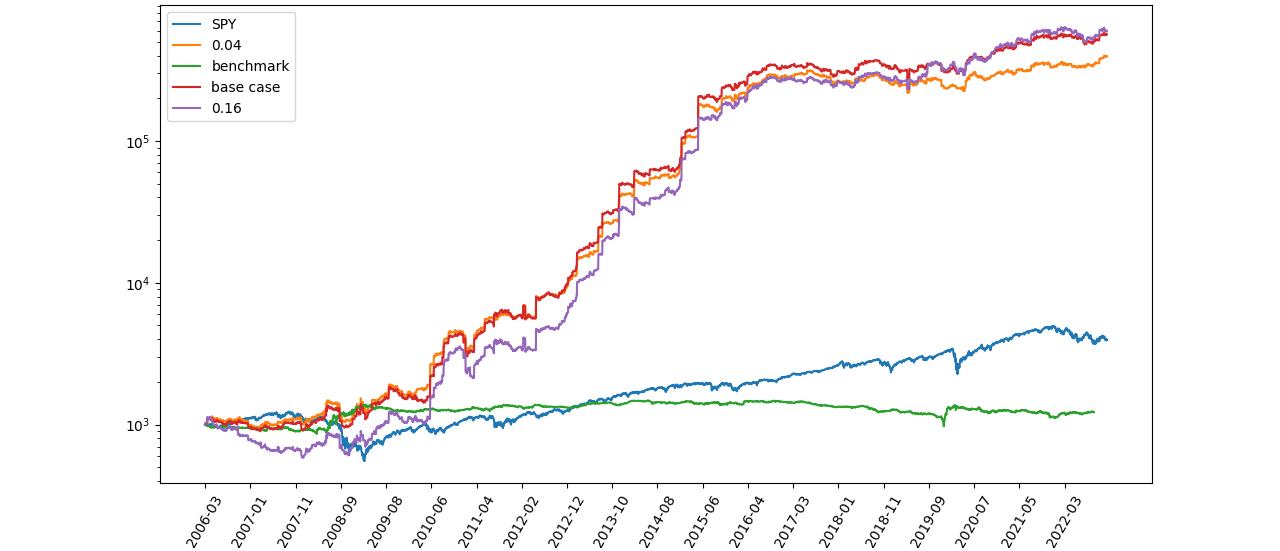}
        \subcaption*{\footnotesize{Source: Own Elaboration. Own backtesting implementation, out-of-sample performance between 03.2006 and 12.2022}}
    \end{figure}
\subsection{Threshold applied in construction of the in-sample dataset}
\par The next tested parameter is the threshold applied in the construction of the in-sample dataset. Namely, the take profit threshold required for the signal to be considered the representative of the class of profitable signals. The change of this parameter has a substantial effect on the strategy performance (Figure 15). The values of IR* have decreased significantly when compared to the base case (Table 16). The MDD has more than doubled for the strategy with the parameter equal to 0.08. Thus, the strategy is sensitive to changes in this parameter.
\begin{table}[H]
    \centering
    \caption{Performance metrics for sensitivity analysis of threshold applied in the construction of in-sample dataset}
    \label{table:1}
    \begin{tabular}{c c c c c c} 
    \hline
    Metric & Base Case & Benchmark &  0.02 &  0.08 & SPY \\ [0.5ex] 
    \hline
    ARC & \textbf{49.33\% }& 1.13\% & 34.47\% & 30.57\% & 9.12\%  \\ 
    ASD & 38.01\$ & \textbf{9.16\%} & 37.88\% & 43.33\% & 20.11\% \\
    IR* & \textbf{1.30} & 0.14 &  0.91 & 0.71 & 0.45 \\
    SORTINO & \textbf{3.38} & 0.29 & 2.05 & 1.32 & 0.70 \\
    MDD & \textbf{31.98\%} & 34.30\% & 45.70\% & 68.40\% &  55.19\% \\ 
    MLD & \textbf{2.10 years} & 3.56 years & 3.72 years & 2.92 years & 4.85 years \\ 
    CR & \textbf{1.54} & 0.04 & 0.75 & 0.45 & 0.17 \\
    IR** &\textbf{ 2.00} & 0.01  & 0.68 & 0.32 & 0.08 \\[1ex] 
    \hline
    \end{tabular}
    \subcaption*{\footnotesize{Source: Own Elaboration. Own backtesting implementation, out-of-sample performance between 03.2006 and 12.2022.\textbf{Base Case}: proposed approach, with weighted ensemble classifier, stop loss and take profit function modifications and kelly criterion. \textbf{Benchmark}: Own implementation of the strategy proposed by Cartea et al. (2023), with 0.05\% transaction costs. \textbf{0.02}: Base Case with halved threshold used in the construction of in-sample dataset. \textbf{0.08}: Base Case with doubled threshold used in the construction of in-sample dataset. \textbf{SPY}: SPY ETF. Bolded values indicate the best metric of all presented strategies.} }
    \end{table}
    \begin{figure}[H]
        \centering
        \caption{Equity curves for sensitivity analysis of threshold applied in the construction of in-sample dataset}
        \label{fig:enter-label}
        \includegraphics[width=1\linewidth]{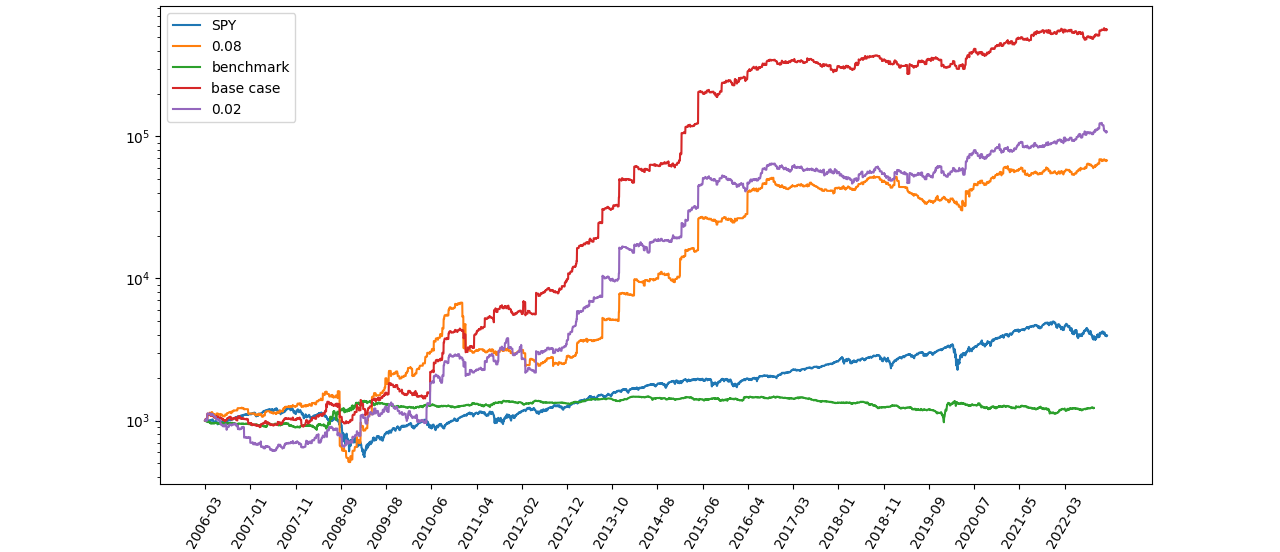}
        \subcaption*{\footnotesize{Source: Own Elaboration. Own backtesting implementation, out-of-sample performance between 03.2006 and 12.2022}}
    \end{figure}
\subsection{Rebalance frequency and length of clustering lookback window}
\par{In order to examine the quality of assumption made in the section 3.3, the sensitivity to changes of frequency of clustering and number of days used to construct correlation matrix which serves as an input to SPONGE$_{sym}$ method. The strategies tested are named according to the convention a\&b where a denotes frequency of rebalance and b the length of lookback window used to construct correlation matrix for clustering purposes. To mitigate the risk of serious classes disbalance in classifers training, threshold used in the construction of in-sample dataset has been scaled linearly by a/10 when a was not equal to 10.}
\begin{table}[H]
    \centering
    \caption{Performance metrics for sensitivity analysis of rebalance frequency and length of clustering lookback window}
    \label{table:1}
    \begin{tabular}{c c c c c c c c c} 
    \hline
    Metric &Base Case& Benchmark &  3\&5 &  10\&5  & 15\&30 & 10\&45& SPY\\ [0.5ex] 
    \hline
    ARC & \textbf{49.33\%} & 1.13\% & -100\% & 16.78\%  & 9.54\% &37.34\% & 9.12\% \\ 
    ASD & 38.01\$ & \textbf{9.16\% }& 35.11\% & 33.48\%  & 46.83\% &59.92\% & 20.11\%\\
    IR* & \textbf{1.30} & 0.14 &  -2.85 & 0.50  & 0.20 & 0.62& 0.45\\
    SORTINO&\textbf{3.38} & 0.29 & -3.18 & 0.94 & 0.95 &1.42 & 0.70 \\
    MDD & \textbf{31.98\%} & 34.30\% & 100\% & 54.02\%  & 37.67\%  &67.02\% &  55.19\%\\ 
    MLD &\textbf{2.10 yrs.}&3.56 yrs.&11.17 yrs.&2.92 yrs.& 4.99 yrs.&4.79 yrs.& 4.85 yrs.\\ 
    CR & \textbf{1.54} & 0.04 & -1 & 0.31  & 0.25 &0.58& 0.17\\
    IR** & \textbf{2.00} & 0.01  & -2.85 & 0.16  & 0.05 & 0.35& 0.08\\[1ex] 
    \hline
    \end{tabular}
    \subcaption*{\footnotesize{Source: Own Elaboration. Own backtesting implementation, out-of-sample performance between 03.2006 and 12.2022 \textbf{Base Case}: proposed approach, with weighted ensemble classifier, stop loss and take profit function modifications and kelly citerion. The values of a and b equal to 10\&30. \textbf{Benchmark}: Own implementation of the strategy proposed by Cartea et al. (2023), with 0.05\% transaction costs.\textbf{SPY}: SPY ETF. Bolded values indicate the best metric of all presented strategies.} }
    \end{table}
    \par{The proposed strategy is very sensitive to changes in both the frequency of portfolio rebalance as well as the length of the lookback window used for clustering. Changes of both parameters in either direction worsened the strategy performance. Changes of clustering frequency seem to have a stronger effect, moreover decreasing the value of both parameters leads to stronger impairment of the IR*, IR**, CR, and Sortino ratio, than increasing them (Table 17). This could be caused by the presence of average cluster returns of last \textbf{10} days as well as average stock returns of last \textbf{10} days in the training dataset, which is highly likely not optimal for strategies tested in this part of the sensitive analysis, especially the one with rebalance every three days. Thus, any attempt at implementation of the proposed strategy, but with different values of the two parameters should also involve redefining features used in the training of the machine learning classifiers. Figure 16 shows, the strongly varying performance of analyzed strategies, and the substandard performance of the 3\&5 strategy, which led to the bankruptcy of that portfolio. The performance of the strategy was most likely caused by classifiers learning noise, due to the issues with incompatible features, described before, which resulted in a lack of ability to trade high-risk stocks like EP or CPWR}
    \begin{figure}[H]
        \centering
        \caption{Equity curves for sensitivity analysis of rebalance frequency and length of clustering lookback window}
        \label{fig:enter-label}
        \includegraphics[width=1\linewidth]{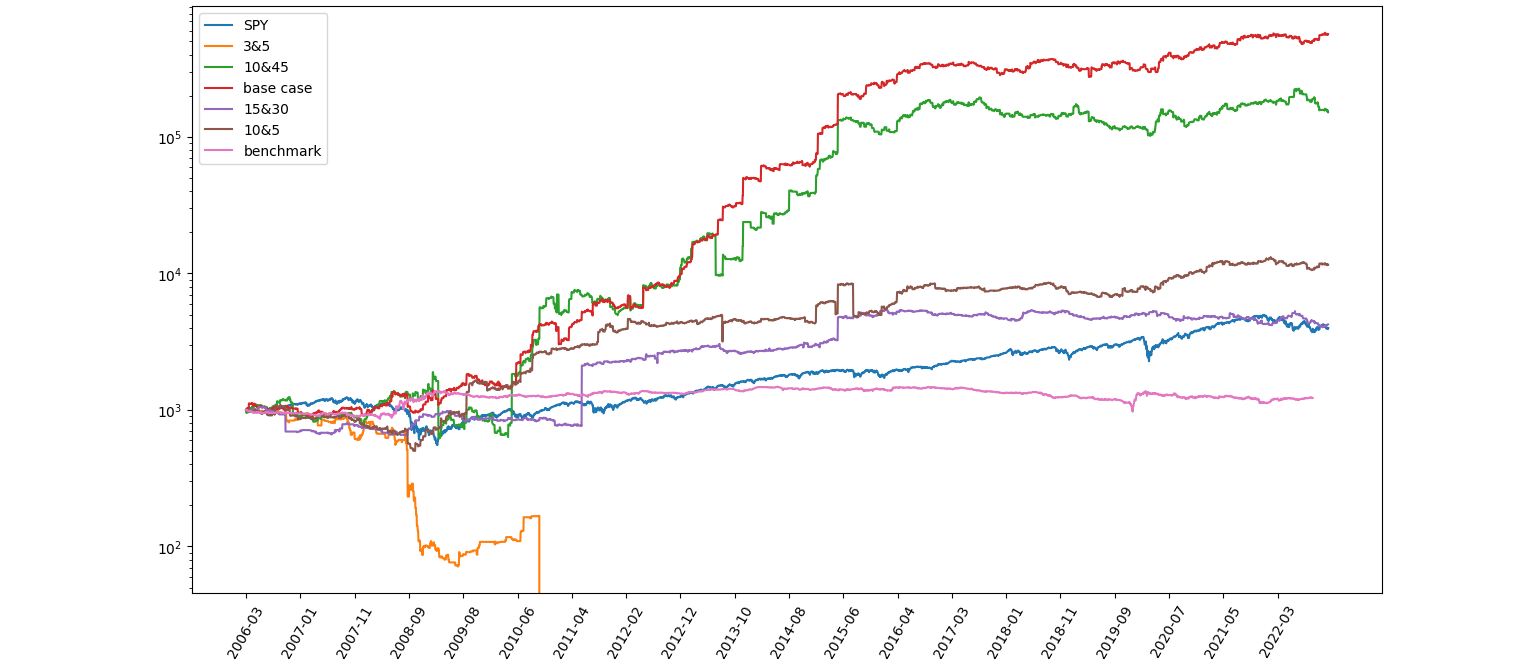}
        \subcaption*{\footnotesize{Source: Own Elaboration. Own backtesting implementation, out-of-sample performance between 03.2006 and 12.2022}}
    \end{figure}
\subsection{Summary of the sensitivity analysis}
\par{In order to summarize the sensitivity analysis, Table 18 has been provided. The table shows the number of performance metrics on which a given strategy performed best within the sensitivity analysis of a given parameter. Overall, in case of most of the parameters, the base case strategy has been the best on the majority of the metrics, indicating substantial sensitivity of the selected} approach. 
\begin{table}[H]
\centering
\caption{Summary of the sensitivity analysis conducted}
\label{tab:my_label}
\begin{tabular}{c c c c }
\hline
\multicolumn{2}{c}{\textbf{Stop loss and take profit modifications}} & \multicolumn{2}{c}{\textbf{Take Profit Threshold}} \\
\hline
No time-variant, no Kelly &  1     & 0.04 & 2 \\
No time Variant, Kelly & 2   & 0.08 (Base Case)& 5     \\
Time variant \& Kelly (Base Case) &5  & 0.16 &  1 \\
\hline
\multicolumn{2}{c}{\textbf{Ensemble weights}} & \multicolumn{2}{c}{\textbf{\makecell{Threshold for constructing\\ in-sample dataset}}} \\
\hline
Equally Weighted    & 0 & 0.02     & 1\\
HistGradientBoosting weight doubled (Base Case)& 7  & 0.04 (Base Case)  & 7\\
Only HistGradientBoosting& 1  & 0.08& 0    \\
\hline
\multicolumn{2}{c}{\textbf{Transaction costs}} & \multicolumn{2}{c}{\textbf{Stop loss threshold}} \\
\hline
0\%& 7      &  1\%    & 1\\
0.05\% (Base Case)& 1      & 3\%      & 1\\
0.075\% &  0     &   5\% (Base Case)    & 6\\
0.1\% &   0    & 10\%     & 0\\
\hline
\multicolumn{2}{c}{\textbf{\makecell{Rebalance frequency and length\\ of clustering look back window}}} & \multicolumn{2}{c}{\textbf{Overperforming stocks exclusion}} \\
\hline
3\&5&   1    &  With EP and CPWR (Base Case)     & 5\\
10\&5&   0    &  Without  EP and CPWR    & 1\\
10\&30 (Base Case) & 7      &  Without EP     & 2\\
10\&45&   0    &    Without CPWR   & 0\\
15\&30&    0   &       &\\
\hline
\end{tabular}
\subcaption*{\footnotesize{Souce: Own elaboration. The table shows the number of performance metrics on which a given strategy performed the best within the sensitivity analysis of a given parameter.}}
\end{table}

\section{Conclusions and Further Research}
\par The study examined the performance of the algorithmic trading strategy, in which the ensemble of machine learning classifiers has been applied in order to filter the signals generated by a statistical arbitrage framework based on graph clustering. The work was based on the price data of historical constituents of the S\&P500 index, downloaded from the Yahoo finance database. Cartea et al. (2023) strategy has been modified by decreasing the trading frequency. Subsequently, an In-sample dataset was created from the first 1500 trading days, which contained information about signals generated by that strategy and their performance. That dataset has been split into training and validation fractions. The training fraction has been utilized to train and find optimal hyperparameters of five machine learning classifiers, including Multi-layer perceptron, Histogram Gradient Boosting, Ada Boost, Logistic regression, and stochastic gradient descent. Their performance has been assessed on the validation fraction of the dataset, using the Brier and Precision scores.  Finally, the strategy in which signals were filtered using the classifier has been tested on the out-of-sample part of the dataset. Several optimizations, including, the Kelly criterion, novel time-variant stop loss and take profit functions, and likelihood-weighed positions have been added and tested to accurately assess the potential of the graph-based methods. A thorough evaluation of the strategy performance, measured by various metrics and deepened in the sensitivity analysis, has been performed. The obtained results suggest the validity of the approach, as all of the variants of the classifier-based strategy outperformed both the original graph clustering-based statistical arbitrage strategy and the index-tracking ETF buy-and-hold approach. These results have been obtained under realistic transaction cost assumptions. Thus, the following answers could be given to the formulated research questions:
\begin{itemize}
    \item \textit{RQ1: Does the usage of signal quality classifiers improve the quality of the graph-clustering-based statistical arbitrage strategy?} - The results of the backtesting, suggest that the usage of the machine learning classifiers can indeed improve the performance of the strategy. The conclusion is supported by the higher values of Calmar, Sortino, and modified information ratios obtained by various versions of the strategy which used the classifiers when compared to the one that didn't. The improvement seemed to stem from the successful reduction of the volume of unprofitable signals traded. The best results have been observed when using the weighted soft voting ensemble of classifiers. 
    \item \textit{RQ2: To what extent, does the implementation of transaction and risk management measures to influence the performance of the strategy?} - After performing the sensitivity analysis, in which the impact of removing the implemented optimization has been measured, it can be concluded that their influence is limited. The addition of neither the Kelly criterion nor stop loss and take profit function modification has led to an increase in the Information Ratio*. However, the latter seems to have a moderate positive effect on downside and tail risk reduction. Thus the novel approach of multiplying the stop functions by the predicted probability of the signal's profitability as well as the time-variant factor seems to have potential for further examination. The very limited impact of the usage of the Kelly criterion is most likely caused by the very rudimentary and simple implementation.
    \item \textit{RQ3: What is the sensitivity of the strategy to changes in transaction costs?} - Sensitivity analysis has clearly demonstrated that the changes in transaction costs have a limited impact on strategy performance. Strategy can be considered robust to changes of that parameter, as doubling it resulted in only a 10\% drop of Information Ratio*.
\item \textit{RQ4: To what extent does the change of weights in the classifiers ensemble influence the strategy} - Changes of weights used in the construction of the ensemble have a measurable impact on the performance of the strategy. Sensitivity analysis suggests, that both equally weighted ensemble and selection of only the best classifier lead to results that are worse than weighted ensemble with the increased weight of the best-performing model.
\end{itemize}
\par It needs to be acknowledged, that the study is constrained by certain limitations. Firstly, the data quality in the Yahoo finance database is certainly not perfect. Although the reproducing of the methodology of Cartea et. al. (2023) resulted in very similar values of the performance metrics, suggesting that the obtained results are representative, the extent of missing data has been noticeable, thus increasing the uncertainty of the backtesting results. Moreover, no slippage effects have been considered in the backtesting, which could also be a liberal assumption, given that the strategy profited significantly from price movements of highly volatile stocks. Moreover, the precision and negative brier scores of the classifiers on the validation datasets have not been very high, which was likely caused by not optimal feature engineering and the lack of feature elimination. 
\par After conducting the study, a few possible ideas for further research can be identified. Firstly, the drastic change in the strategy's performance induced by the elimination of the overperforming, volatile stocks may suggest that the strategy can potentially perform well in environments in which assets commonly possess such characteristics. An example of a market worth considering could be the cryptocurrency market. Moreover, it needs to be noted that the overall design of the classifiers has been rather simple. A further investigation worth conducting would test how the performance of the classifiers changes, once the multilabel classification is utilized. For example, signals can be classified as prospective profitable shorts, profitable longs, or the ones that are not likely to be profitable. Of course, the whole framework can also be tested on higher frequencies data, which are commonly used in the statistical arbitrage strategies.

\section{References}

\begin{enumerate}[label={[\arabic*]}]

    \item Bieganowski, B., \& Slepaczuk, R. (2024). Supervised Autoencoder MLP for Financial Time Series Forecasting. arXiv preprint arXiv:2404.01866.
    \item Bui Q., Ślepaczuk R., (2022), Applying Hurst Exponent in Pair Trading Strategies on Nasdaq 100 index, Physica A: Statistical Mechanics and its Applications 592, p.126784, ISSN = 0378-4371, https://doi.org/10.1016/j.physa.2021.126784
    \item Bronakowska, K., Naumowicz, P., \& Ślepaczuk, R. (2021). Various Approaches to Algorithmic Investment Strategies on Equity Indices and Stocks Markets, ArchaeGraph, ISBN: 978-83-67074-34-6
    \item Borovkova, S., \& Tsiamas, I. (2019). An ensemble of LSTM neural networks for high‐frequency stock market classification. Journal of Forecasting, 38(6), 600-619.
    \item Cartea, Á., Jin, Q., \& Cucuringu, M.  (2023). Correlation Matrix Clustering for Statistical Arbitrage Portfolios. In Proceedings of the Fourth ACM International Conference on AI in Finance (pp. 557-564), Available at SSRN: \\
    https://papers.ssrn.com/sol3/papers.cfm?abstract\_id=4560455
    \item Chen, H., Chen, S., Chen, Z., \& Li, F. (2019). Empirical investigation of an equity pairs trading strategy. Management Science, 65(1), 370-389.
    \item Chen, L., Sun, L., Chen, C. M., Wu, M. E., \& Wu, J. M. T. (2021). Stock trading system based on machine learning and kelly criterion in internet of things. Wireless Communications and Mobile Computing, 2021, 1-9.
    \item Cheng, D., Yang, F., Xiang, S., \& Liu, J. (2022). Financial time series forecasting with multi-modality graph neural network. Pattern Recognition, 121, 108218.
    \item  Cucuringu, M., Davies, P., Glielmo, A., \& Tyagi, H. (2019). SPONGE: A generalized eigenproblem for clustering signed networks. ArXiv, abs/1904.08575.   
    \item Dixon, M., Klabjan, D., \& Bang, J. H. (2017). Classification-based financial markets prediction using deep neural networks. Algorithmic Finance, 6(3-4), 67-77.
    \item Do, B., \& Faff, R. (2011). Are Pairs Trading Profits Robust to Trading Costs?. Journal of Financial Research, 35.
     \item Faff, R., \& Anderson, J. (2004). Maximizing futures returns using fixed fraction asset allocation. Applied Financial Economics, 14, 1067-1073.
    \item Fama, E. F., \& French, K. R. (1997). Industry costs of equity. Journal of financial economics, 43(2), 153-193.
    \item Farrell J. A. "SP500" (2024). available at: https://github.com/fja05680/sp500?tab=MIT-1-ov-file, last accessed 17.04.2024
    \item Freund, Y., \& Schapire, R. E. (1997). A decision-theoretic generalization of on-line learning and an application to boosting. Journal of computer and system sciences, 55(1), 119-139.
      \item Gatev, E., Goetzmann, W. N., \& Rouwenhorst, K. G. (2006). Pairs trading: Performance of a relative-value arbitrage rule. The Review of Financial Studies, 19(3), 797-827.
    \item Hastie, T., Tibshirani, R., Friedman, J. H., \& Friedman, J. H. (2009). The elements of statistical learning: data mining, inference, and prediction (Vol. 2, pp. 1-758). New York: springer.
  \item Kelly, J. L. (1956). A new interpretation of information rate. the bell system technical journal, 35(4), 917-926.
   \item Kim, K. (2011). Performance Analysis of Pairs Trading Strategy Utilizing High Frequency Data with an Application to KOSPI 100 Equities. SSRN Electronic Journal.
    \item Kosc, K., Sakowski, P., \& Ślepaczuk, R. (2019). Momentum and contrarian effects on the cryptocurrency market. Physica A: Statistical Mechanics and its Applications, 523, 691-701.
    \item Kryńska, K., \& Ślepaczuk, R. (2023). Daily and intraday application of various architectures of the LSTM model in algorithmic investment strategies on Bitcoin and the S\&P 500 Index. Available at SSRN 4628806.
  \item Leung, M. T., Daouk, H., \& Chen, A. S. (2000). Forecasting stock indices: a comparison of classification and level estimation models. International Journal of forecasting, 16(2), 173-190.
    \item Li, X., Wang, J., Tan, J., Ji, S., \& Jia, H. (2022). A graph neural network-based stock forecasting method utilizing multi-source heterogeneous data fusion. Multimedia tools and applications, 81(30), 43753-43775.
    \item Li, B., Xie, K., Lu, S., Lin, J., \& Wang, Z. (2020, July). LSTM-Based quantitative trading using dynamic K-Top and Kelly criterion. In 2020 International Joint Conference on Neural Networks (IJCNN) (pp. 1-8). IEEE.
  \item Markowitz, H. (1952). Portfolio Selection. The Journal of Finance, 7(1), 77–91. \\
  https://doi.org/10.2307/2975974
  \item Miao, G. J. (2014). High frequency and dynamic pairs trading based on statistical arbitrage using a two-stage correlation and cointegration approach. International Journal of Economics and Finance, 6(3), 96-110.
\item Pedregosa, F., Varoquaux, G., Gramfort, A., Michel, V., Thirion, B., Grisel, O., Blondel, M., Prettenhofer, P., Weiss, R., Dubourg, V., \& others (2011). Scikit-learn: Machine learning in Python. the Journal of machine Learning research, 12, 2825–2830.
\item Statman, M. (1987). How Many Stocks Make a Diversified Portfolio? The Journal of Financial and Quantitative Analysis, 22(3), 353–363. https://doi.org/10.2307/2330969
    \item Stübinger, J., \& Bredthauer, J. (2017). Statistical arbitrage pairs trading with high-frequency data. International Journal of Economics and Financial Issues, 7(4), 650-662.
  \item Thorp, E. O. (1969). Optimal gambling systems for favorable games. Revue de l'Institut International de Statistique, 273-293.
  \item Worasucheep, C. (2022). Ensemble classifier for stock trading recommendation. Applied Artificial Intelligence, 36(1), 2001178.
  \item Yangru, W., \& Zhang, H. (1997). Forward premiums as unbiased predictors of future currency depreciation: A non-parametric analysis. Journal of International Money and Finance, 16(4), 609-623.
  \item Yin, T., Liu, C., Ding, F., Feng, Z., Yuan, B., \& Zhang, N. (2022). Graph-based stock correlation and prediction for high-frequency trading systems. Pattern Recognition, 122, 108209.
    \item Zawadowski M., (2020), Wykład ze Wstpu do Informatyki (Jezyk C), Available at: https://tinyurl.com/2526ct5v
\item Zelenkov, Y. (2017). Parallel heterogeneous multi-classifier system for decision making in algorithmic trading. In Supercomputing: Third Russian Supercomputing Days, RuSCDays 2017, Moscow, Russia, September 25–26, 2017, Revised Selected Papers 3 (pp. 251-265). Springer International Publishing.
  \item Zhan, H. C. J., Rea, W., \& Rea, A. (2015). An application of correlation clustering to portfolio diversification. arXiv preprint arXiv:1511.07945.

\end{enumerate}

\end{document}